\documentclass[10pt,final,doublecolumn]{IEEEtran}
\usepackage{amsmath}
\usepackage{latexsym}
\usepackage{graphicx}
\usepackage{bbding}
\usepackage{indentfirst}
\usepackage{algorithm,algorithmic}
\usepackage{setspace}
\usepackage{float}
\usepackage{epstopdf}
\usepackage{amssymb}
\usepackage{amsfonts}
\usepackage{enumerate}
\usepackage{multicol}
\usepackage{color}
\usepackage{slashbox}
\usepackage{bm}
\usepackage{amssymb}
\usepackage{stfloats}
\usepackage{epstopdf}
\usepackage{threeparttable}
\usepackage{epstopdf}
\usepackage{threeparttable}
\usepackage{subfigure}
\usepackage{makecell}

\usepackage{cite}
\IEEEoverridecommandlockouts
\allowdisplaybreaks[4]

\begin{document}
\title{6D Movable Antenna Enhanced Wireless Network Via Discrete Position and Rotation Optimization}

\author{{Xiaodan Shao,~\IEEEmembership{Member,~IEEE}, Rui Zhang, \IEEEmembership{Fellow, IEEE}, Qijun Jiang, and Robert
Schober,~\IEEEmembership{Fellow, IEEE}}

\thanks{X. Shao is with the Institute for Digital Communications, Friedrich-Alexander-University Erlangen-Nurnberg (FAU), 91054
Erlangen, Germany (email:xiaodan.shao@fau.de).}	

\thanks{R. Zhang is with School of Science and Engineering, Shenzhen Research Institute of Big Data, The Chinese University of Hong Kong, Shenzhen, Guangdong 518172, China. He is also with the Department of Electrical and Computer Engineering, National University of Singapore, Singapore 117583 (e-mail: elezhang@nus.edu.sg).}

\thanks{Q. Jiang is with the School of Science and Engineering, Chinese University of Hong Kong, Shenzhen, China 518172 (e-mail:
qijunjiang@link.cuhk.edu.cn).}

\thanks{R. Schober is with the Institute for Digital Communications, Friedrich-Alexander-University Erlangen-Nurnberg (FAU), 91054
Erlangen, Germany (email: robert.schober@fau.de).}

}
\maketitle

\IEEEpeerreviewmaketitle

\begin{abstract}
Six-dimensional movable antenna (6DMA) is an effective approach to improve wireless network capacity by adjusting the 3D positions and 3D rotations of distributed antenna surfaces based on the users' spatial distribution and statistical channel information. Although continuously positioning/rotating 6DMA surfaces can achieve the greatest flexibility and thus the highest capacity improvement, it is difficult to implement due to the discrete movement constraints of practical stepper motors. Thus, in this paper, we consider a 6DMA-aided base station (BS) with only a finite number of possible discrete positions and rotations for the 6DMA surfaces. We aim to maximize the average network capacity for random numbers of users at random locations by jointly optimizing the 3D positions and 3D rotations of multiple 6DMA surfaces at the BS subject to discrete movement constraints. In particular, we consider the practical cases with and without statistical channel knowledge of the users, and propose corresponding offline and online optimization algorithms, by leveraging the Monte Carlo and conditional sample mean (CSM) methods, respectively.
Simulation results verify the effectiveness of our proposed offline and online algorithms for discrete position/rotation optimization of 6DMA surfaces as compared to various benchmark schemes with fixed-position antennas (FPAs) and 6DMAs with limited movability. It is shown that 6DMA-BS can significantly enhance wireless network capacity, even under discrete position/rotation constraints, by exploiting the spatial distribution characteristics of the users.
\end{abstract}

\begin{IEEEkeywords}
6D movable antenna, discrete position and rotation optimization, base station (BS) architecture, non-convex integer programming, conditional sample mean (CSM).
\end{IEEEkeywords}

\section{Introduction}
Future wireless networks, such as the six-generation (6G) and beyond, need to support communications for increasingly more densely distributed wireless devices, while incurring
low power consumption and ensuring super reliability. To achieve this challenging goal, the current technology trend is to equip the base station (BS) or the entire wireless network with a rapidly growing number of antennas, e.g., from massive multiple-input multiple-output (MIMO) \cite{LA10,ruig} to cell-free massive MIMO \cite{free,free0}, and extremely large-scale MIMO \cite{exl,zeng}.
However, this approach inevitably results in steadily increasing hardware cost, energy consumption, and signal processing/computational complexity, which thus cannot fulfill the high performance and efficiency expectations of future wireless networks completely. One major limitation of current MIMO communication systems lies in the fact that the antennas are deployed at fixed positions at the BS or distributed access points. As a result, given that the total number of antennas is fixed, the wireless network cannot allocate its antenna resources flexibly based on the spatially non-uniform distributions of the users, beyond the traditional adaptive MIMO processing (e.g., transmit precoding, receive combining).
Thus, with fixed-position antennas (FPAs), the adaptation to changes in the 3D spatial distribution of users is very limited.
\begin{figure}[t!]
\centering
\setlength{\abovecaptionskip}{0.cm}
\includegraphics[width=3.2in]{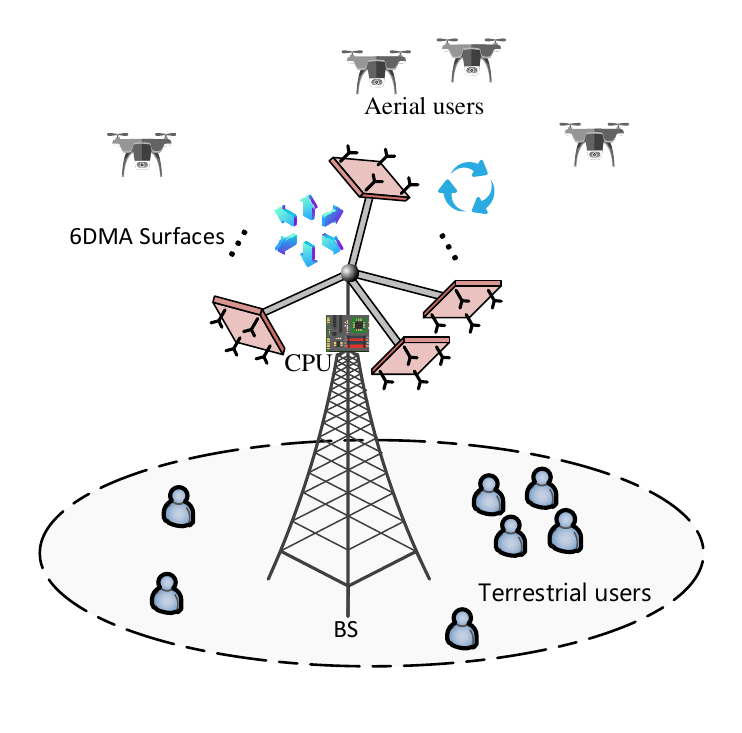}
\caption{6DMA-enabled BS for adapting to non-uniform user distributions in wireless networks.}
\label{practical_scenario}
\end{figure}

Recently, six-dimensional movable antenna (6DMA) has been proposed as a new and
effective solution for improving MIMO system capacity without the need of adding more antennas, by fully exploiting the adaptability of the 3D positions and 3D rotations of limited number of antennas at the BS \cite{first}. As shown in Fig. \ref{practical_scenario}, a 6DMA-enabled BS generally comprises
a number of 6DMA surfaces each of which can be independently adjusted in terms of both 3D position and 3D rotation, subject to practical movement constraints.
Each 6DMA surface is connected to a central processing unit (CPU) at the BS via an extendable and rotatable rod, which contains flexible wires for providing power supply to the 6DMA surface and facilitating radio-frequency(RF)/control signal exchange between it and the CPU. At the two ends of the rod, two motors are installed which are controlled by the CPU to adjust the 3D position and 3D rotation of the connected 6DMA surface.
By jointly designing and tuning the positions and rotations of all 6DMA surfaces
adaptively according to the users' spatial distribution (and the corresponding long-term/statistical channel state information (CSI)), it was shown in \cite{first} that 6DMA-BSs can significantly improve the network capacity compared to conventional BSs with FPAs, with only slow/infrequent adjustment of the positions/rotations of 6DMA surfaces. The appealing capacity gain is mainly due to the adaptive positioning/rotating of the antennas at the BS allocating them to match the spatial user distribution, which enhances not only the channel gains by fully exploiting directionality of antennas, but also the spatial multiplexing gains and interference mitigation capabilities, especially when the user distribution is spatially non-uniform.

We note that the 6DMA systems considered in \cite{first} and in this paper differ significantly from the existing
fluid antenna (FA) system \cite{9388928,9264694}, or 2D movable antenna (2DMA) systems \cite{zhu2, ma, yifei}, and
has unique advantages in comparison with them.
Firstly, the existing works on FA/2DMA predominantly consider the movement of antennas along a given line or on a given 2D surface with a finite length/area. As a result, a BS equipped with FA/2DMA
is unable to adapt to the variations of the 3D spatial user distribution (see Fig. \ref{practical_scenario}) due to their limited degrees of freedom (DoFs) in movability.
In contrast, the proposed 6DMA-BS can more flexibly adjust the 3D positions and 3D rotations of antennas/antenna surfaces via mechanical motors, and therefore can more effectively adapt to spatially non-uniform user distribution/channel statistics.
Secondly, in existing works on FA/2DMA, the individual antennas usually serve as the basic movable units to maximally exploit the instantaneous channel spatial variation, which requires frequent antenna movement and thus incurs high implementation cost and time overhead, especially for applications with fast time-varying channels.
In contrast, the proposed 6DMA-BS moves the entire antenna surface as a basic  unit, which reduces implementation cost and complexity. In addition, 6DMA surfaces need to be adjusted in position and rotation much less frequently as the users' spatial distribution and channel statistics are expected to vary only slowly over
time.

In \cite{first}, the highly optimistic assumption is made that 3D positions and 3D rotations of all 6DMA surfaces can be adjusted in a continuous manner, so as to exploit the greatest flexibility in movement and thus achieve the highest capacity gain. However, in practice, this is difficult to realize since 6DMA surfaces need to be mechanically moved by physical devices, such as a stepper motor, which can only adjust the position/rotation of each 6DMA surface in discrete steps, which are usually specified by the motor used. As such, discrete position/rotation adjustments of 6DMA surfaces will limit their spatial DoFs for adaption and thus inevitably cause capacity degradation as compared to their continuous counterparts.
Hence, for achieving a network capacity gain with the 6DMA-BS, the joint optimization of the 3D positions and 3D rotations of the 6DMA surfaces becomes more crucial in the discrete case, which thus motivates the current work.

In this paper, we study a new 6DMA system, where the 6DMA-BS is equipped with a number of finite-size antenna surfaces (e.g., uniform planar array (UPA)), which can be independently adjusted in terms of both 3D position and 3D rotation each to be selected from a predefined set of discrete values. Different from the existing works on FA/2DMA (e.g., \cite{9388928,9264694,zhu2, ma, yifei}), which either assume known instantaneous CSI or employ channel estimation for antenna movement optimization, we optimize the 3D position/rotation of 6DMA surfaces based on only statistical channel knowledge of the users in the network. This knowledge may be a priori available (e.g., since user channel data was well collected over a long time period) or a priori unavailable (e.g., because BS has been newly deployed or the user distribution in an area changes abruptly). Therefore, in this paper, we optimize the 3D positions and 3D rotations of 6DMA surfaces for maximization of the average network capacity for the cases with and without the statistical channel knowledge, respectively. The main contributions of this paper are summarized as follows.
\begin{itemize}
\item
First, by taking into account the practical limitations on discrete 3D positions and 3D rotations of 6DMA surfaces, we extend the 6DMA-BS system model,  continuous antenna movement constraints, and  channel model in \cite{first}, to the new setup with only discrete position/rotation adjustment.
Then, we formulate a new optimization problem for maximization of the average network capacity by jointly selecting the 3D positions and 3D rotations of the 6DMA surfaces, subject to their discrete movement constraints at the BS.

\item
Next, we propose an offline optimization method to solve the formulated problem assuming {\it a priori}  knowledge of channel statistics of users in the network. Specifically, we first apply the Monte Carlo method to generate a set of channel realizations independently based on the statistical channel knowledge and approximate the network capacity as the average of the users' achievable sum-rates over these channel realizations. The resulting optimization problem is a non-convex integer program, which is difficult to solve optimally. Thus, we develop a series of problem reformulations to convert the original problem to a tractable form with a convex feasibility set after relaxing the integer variables to continuous ones. Then, the problem is efficiently solved via the conditional gradient method \cite{linear, mk}.

\item
Furthermore, we propose an online algorithm for the optimization of the positions and rotations of 6DMA surfaces without any prior knowledge of the channel statistics. To this end, based on the Fibonacci Sphere, we first propose a new method for generating the discrete positions and rotations of the 6DMA surfaces \cite{trip}, which always satisfy the practical surface movement constraints over the considered discrete set. Based on this, we then propose a new online optimization method by randomly generating multiple sets of combinations of discrete positions and rotations for all 6DMA surfaces and measuring the actual achievable sum-rate of the users for each set. With such measured data acquired, we can optimize the 3D positions/rotations of all available 6DMA surfaces directly for maximization of the average network capacity, without the need for explicit estimation of statistical channel knowledge. This is achieved by applying a low-complexity conditional sample mean (CSM) method, which has also been recently applied to solve the optimization of massive reflection coefficients for intelligent reflecting surface (IRS) aided communication systems without CSI \cite{proc, shuyi, shaoirs}.

\item
Finally, we evaluate the performance of our proposed
offline/online algorithms for jointly optimizing the 3D positions and 3D rotations of the 6DMA surfaces with/without statistical channel knowledge via simulations. The results demonstrate that the proposed 6DMA-BS can significantly improve the network capacity over benchmark BS architectures employing FPAs or 6DMAs with limited movability, even under discrete position/rotation constraints, by exploiting the characteristics of the spatial distribution of the users.
\end{itemize}
\begin{figure}[t!]
\centering
\setlength{\abovecaptionskip}{0.cm}
\includegraphics[width=3.39in]{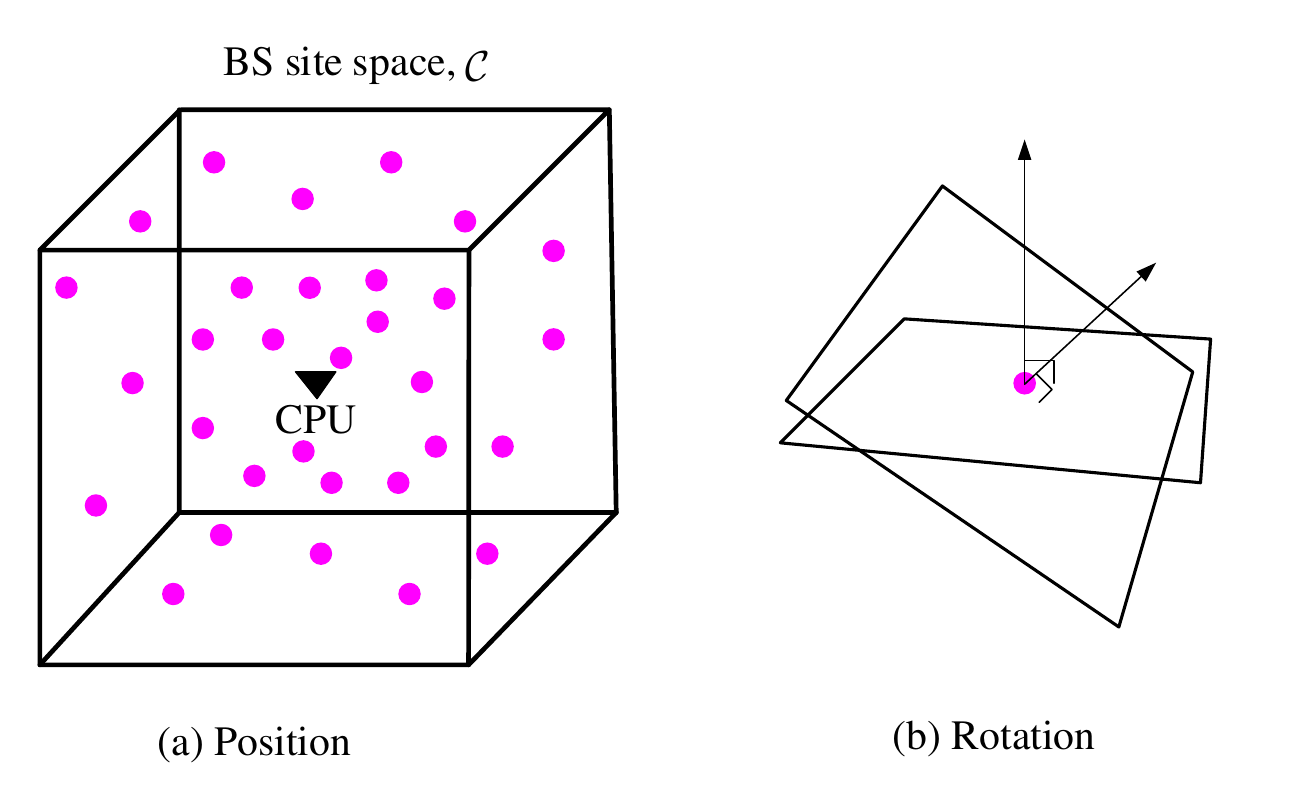}
\caption{Discrete positions and rotations of 6DMA surfaces (circles represent different positions).}
\label{position_rotation}
\end{figure}

The rest of this paper is organized as follows. Section II presents the 6DMA-BS system model with discrete 3D position/rotation adjustment of 6DMA surfaces and the corresponding practical movement constraints and channel model.
In Section III, we formulate the optimization problem for selecting the discrete 3D positions and 3D rotations of all 6DMA surfaces to maximize the network capacity.
In Sections IV and V, we present our proposed offline and online algorithms to solve the formulated problem with and without statistical knowledge of users' channels, respectively. Section VI provides numerical results and pertinent discussions. Finally, Section VII concludes this paper.

\emph{Notations}: Boldface upper-case and lower-case letters denote
matrices and vectors, respectively, $(\cdot)^H$ and $(\cdot)^T$ respectively denote conjugate transpose and transpose, $\mathbb{E}[\cdot]$ denotes the statistical expectation,  $\hat{\mathbb{E}}[\cdot]$ denotes the sample mean, $\left \|\cdot\right \|_2$ denotes the Euclidean norm, $\mathbf{0}_{N}$ denotes the $N\times 1$ vector with all zero elements, $\mathbf{I}_N$ denotes the $N\times N$ identity matrix, $\mathrm{diag}({\bf x})$ denotes a diagonal matrix with the diagonal entries specified by vector ${\bf x}$, $|\mathcal{X}|_{\mathrm{c}}$ denotes the cardinality of set $\mathcal{X}$, $\wedge$ denotes cross product between two vectors, $\mathrm{vec}(\mathbf{A})$ denotes the stacking of the columns of matrix $\mathbf{A}$ one by one in a single column, $[\mathbf{A}]_{i,:}$ denotes the $i$-th row of matrix $\mathbf{A}$, $[\mathbf{a}]_j$ denotes the $j$-th element of vector $\mathbf{a}$, $[\mathbf{A}]_{i,j}$ denotes
the element of matrix $\mathbf{A}$ in the $i$-th row and $j$-th column,  $\lceil\cdot\rceil$ denotes the ceiling operator, $\mathcal{O}(\cdot)$ denotes the big-O notation, $\max\{ \cdot\}$ and $\min\{\cdot\}$ denote the selection of the maximum and minimum values from a given set, respectively, $\arctan2(\cdot)$ is the two argument arctangent function, and $\cup$ denotes the union of two sets.

\section{System Model}
\subsection{6DMA-BS Model}
As shown in Fig. \ref{practical_scenario}, we consider
a 6DMA-BS equipped with $B$ 6DMA surfaces. Each 6DMA surface is modeled as a UPA with $N\geq 1$ antennas.
These 6DMA surfaces are connected to a CPU at the BS, which can provide power supply to the 6DMA surfaces, exchange signals with them, and control their 3D positions and 3D rotations.

\begin{figure}[t!]
\flushleft
\setlength{\abovecaptionskip}{0.cm}
\includegraphics[width=3.6in]{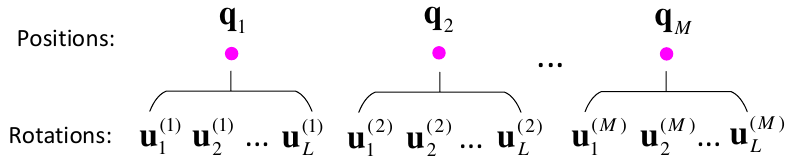}
\caption{Relationship between discrete positions and rotations for 6DMA surfaces.}
\label{MCSM}
\end{figure}

\begin{figure}[t]
\centering
\setlength{\abovecaptionskip}{0.cm}
\includegraphics[width=3.51in]{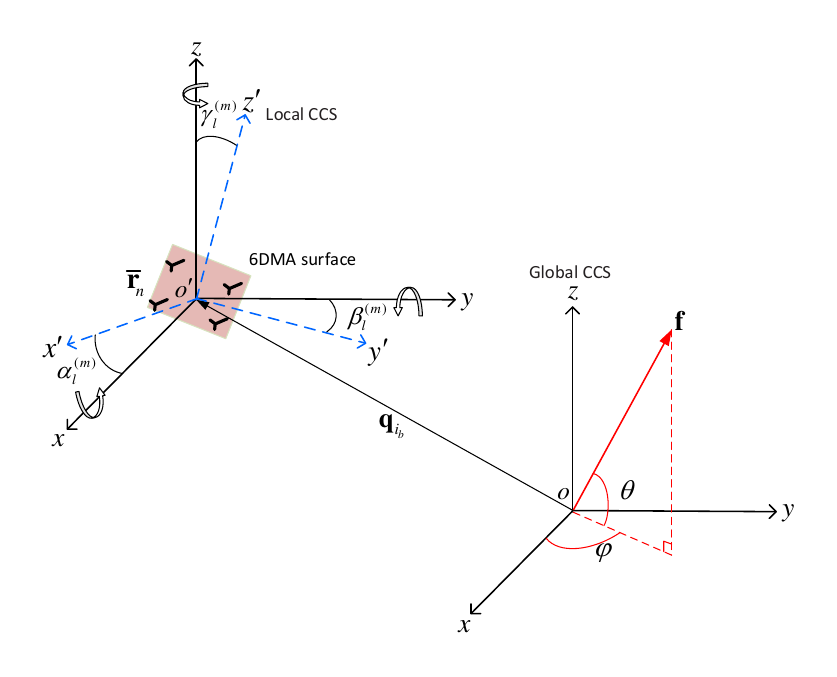}
\caption{Illustration of the geometry of the 6DMA surface at the $m$-th position.}
\label{system}
\end{figure}

For ease of practical implementation, as shown in Fig. \ref{position_rotation}, we assume that the position/rotation of each 6DMA surface can take only a finite number of discrete values. Specifically, we assume that there are in total $M\geq B$ discrete
positions, denoted by set $\mathcal{M}=\{1,2,\cdots,M\}$, and $L\geq 1$ discrete rotations at each position, denoted by set $\mathcal{L}=\{1,2,\cdots,L\}$. As shown in Fig. \ref{MCSM}, we denote $\mathbf{q}_m\in \mathbb{R}^3, m\in \mathcal{M}$, as the $m$-th discrete  position, and $\mathbf{u}_{l}^{(m)}\in \mathbb{R}^3, l\in \mathcal{L}$, as the $l$-th discrete rotation at the $m$-th discrete position. Specifically, they are respectively given by
\begin{align}\label{bb}
\mathbf{q}_{m}&=[x_{m},y
_{m},z_{m}]^T\in\mathcal{C},\\
\mathbf{u}_{l}^{(m)}&=[\alpha_{l}^{(m)},\beta_{l}^{(m)},\gamma_{l}^{(m)}]^T,
\end{align}
where $x_{m}$, $y_{m}$, and $z_{m}$ denote the coordinates of the center of the 6DMA surface at the $m$-th discrete position in the global Cartesian coordinate system (CCS), denoted as $o\text{-}xyz$, with the CPU's center serving as the origin $o$; $\alpha_{l}^{(m)}$, $\beta_{l}^{(m)}$, and $\gamma_{l}^{(m)}$ denote the corresponding rotation angles with respect to (w.r.t.) the $x$-axis, $y$-axis, and $z$-axis, respectively (see Fig. \ref{system}); and $\mathcal{C}$ denotes the given 3D space at the BS site (assumed to be a convex set with a finite size) within which the 6DMA surfaces can be positioned and rotated dynamically.

\begin{figure*}
\begin{align}\label{R}
\mathbf{R}(\mathbf{u}_{l}^{(m)})=\begin{bmatrix}
c_{\alpha_{l}^{(m)}}c_{\gamma_{l}^{(m)}} & c_{\alpha_{l}^{(m)}}s_{\gamma_{l}^{(m)}} & -s_{\alpha_{l}^{(m)}} \\
s_{\beta_{l}^{(m)}}s_{\alpha_{l}^{(m)}}c_{\gamma_{l}^{(m)}}-c_{\beta_{l}^{(m)}}s_{\gamma_{l}^{(m)}} & s_{\beta_{l}^{(m)}}s_{\alpha_{l}^{(m)}}s_{\gamma_{l}^{(m)}}+c_{\beta_{l}^{(m)}}c_{\gamma_{l}^{(m)}} & c_{\alpha_{l}^{(m)}}s_{\beta_{l}^{(m)}} \\
c_{\beta_{l}^{(m)}}s_{\alpha_{l}^{(m)}}c_{\gamma_{l}^{(m)}}+s_{\beta_{l}^{(m)}}s_{\gamma_{l}^{(m)}} & c_{\beta_{l}^{(m)}}s_{\alpha_{l}^{(m)}}s_{\gamma_{l}^{(m)}}-s_{\beta_{l}^{(m)}}c_{\gamma_{l}^{(m)}} &c_{\alpha_{l}^{(m)}}c_{\beta_{l}^{(m)}} \\
\end{bmatrix}.
\end{align}
\end{figure*}

The rotation matrix $\mathbf{R}(\mathbf{u}_{l}^{(m)})$ corresponding to the rotation angles in $\mathbf{u}_{l}^{(m)}$ is defined in \eqref{R} (which is shown on the top of the next page), where $c_{x}=\cos(x)$ and $s_{x}=\sin(x)$ are defined for notational simplicity \cite{rot3}. Furthermore, we consider the local CCS for the 6DMA surface at each position, denoted by $o'\text{-}x'y'z'$, where the surface center serves as the origin $o'$, and the $x'$-axis is oriented along the direction of the normal vector of the 6DMA surface (see Fig. \ref{system}).
Let $\bar{\mathbf{r}}_{n}$ denote the position of the $n$-th antenna of the 6DMA surface in its local CCS. The position of the $n$-th antenna of the 6DMA surface at the $m$-th discrete position in the global CCS, denoted by $\mathbf{r}_{m,n}\in \mathbb{R}^3$, is then determined by rotation $\mathbf{u}_{l}^{(m)}$ and position $\mathbf{q}_{m}$ as
\begin{align}\label{nwq}
\!\!\!\!\mathbf{r}_{m,n}(\mathbf{q}_{m},\mathbf{u}_{l}^{(m)})=\mathbf{q}_{m}+\mathbf{R}
(\mathbf{u}_{l}^{(m)})\bar{\mathbf{r}}_{n},~n\in\mathcal{N},~m \in\mathcal{M},
\end{align}
where $\mathcal{N} = \{1,2,\ldots, N\}$.

We collect the $M$ possible discrete positions of the 6DMA surfaces in set $\mathcal{Q}=\{\mathbf{q}_1, \mathbf{q}_2, \cdots, \mathbf{q}_M\}$, and for each position $m\in\mathcal{M}$, we collect the $L$ possible discrete rotations of the 6DMA surface in set $\mathcal{U}^{(m)}=\{\mathbf{u}_{1}^{(m)}, \mathbf{u}_{2}^{(m)}, \cdots, \mathbf{u}_{L}^{(m)}\}$. Let $i_{b}\in \mathcal{M}$ and $j_{b}\in \mathcal{L}$ denote respectively the indices of the 3D position and the 3D rotation selected for the $b$-th 6DMA surface, $b\in\mathcal{B} = \{1, 2,\ldots, B\}$. In other words, the position of the $b$-th 6DMA surface is $\mathbf{q}_{i_{b}}$, $\mathbf{q}_{i_{b}}\in\mathcal{Q}$, and the rotation of the $b$-th 6DMA surface is $\mathbf{u}_{j_{b}}^{(i_b)}$, $\mathbf{u}_{j_{b}}^{(i_b)}\in\mathcal{U}^{(i_b)}$. For brevity, we rewrite $\mathbf{u}^{(i_b)}_{j_b}$ equivalently as $\mathbf{u}_{j_b}$ for a given $b$ and thus $i_b$ in the sequel without loss of generality.

To facilitate the discrete position and rotation selection for each 6DMA surface, we introduce position indicator vector $\mathbf{s}_b\in\mathbb{R}^{M}$ and rotation indicator vector $\mathbf{g}_b\in\mathbb{R}^{L}$, $b\in\mathcal{B}$, which are respectively given by
\begin{align}\label{3b}
\!\![\mathbf{s}_b]_m= \left\{\begin{matrix}
1, & \text{if}~i_{b}=m,\\
0, &\mathrm{otherwise},
\end{matrix}\right.
\end{align}
and
\begin{align}\label{3bh}
\!\![\mathbf{g}_b]_l= \left\{\begin{matrix}
1, & \text{if}~j_{b}=l,\\
0, &\mathrm{otherwise}.
\end{matrix}\right.
\end{align}
Note that $\mathbf{s}_b$ and $\mathbf{g}_b$ are uniquely determined by $i_b$ and $j_b$, respectively.

Next, we introduce three constraints on the discrete rotation/position selection for the considered 6DMA surfaces, based on their counterparts given in \cite{first} for the case of continuously rotating/positioning 6DMA surfaces.
\subsubsection{Discrete Rotation Constraints to Avoid Signal Reflection}
The 6DMA surfaces must meet the following rotation constraints to avoid mutual signal reflections between any two 6DMA surfaces,
\begin{align}\label{rcc}
\mathbf{n}(\mathbf{u}_{j_b})^T(\mathbf{q}_{i_v}-\mathbf{q}_{i_b})\leq  0,~\forall b ,v\in \mathcal{B}, ~v\neq b,
\end{align}
where $\mathbf{n}(\mathbf{u}_{j_b})=\mathbf{R}(\mathbf{u}_{j_b})\bar{\mathbf{n}}
$ is the outward normal vector of the $b$-th 6DMA surface in the global CCS, with $\bar{\mathbf{n}}$ being its normal vector in the local CCS.
We define an auxiliary matrix $\mathbf{U}\in\mathbb{C}^{M \times L}$ with its elements given by
\begin{align}\label{uij}
[\mathbf{U}]_{m,l}=\mathbf{n}(\mathbf{u}_l^{(m)})^T\mathbf{q}_m, m\in\mathcal{M}, l\in\mathcal{L}.
\end{align}
Then, by combining \eqref{3b}, \eqref{3bh}, and \eqref{uij}, the rotation constraint \eqref{rcc} can be reformulated into the following form in terms of the position and rotation indicator vectors,
\begin{align}\label{redd0}
(\mathbf{s}_v^T-\mathbf{s}_b^T)
\mathbf{U}\mathbf{g}_b\leq 0,~\forall b, v\in\mathcal{B}, v\neq b.
\end{align}

\subsubsection{Discrete Rotation Constraints to Avoid Signal Blockage}
To prevent signal blockage caused by the rotation of a 6DMA surface towards the CPU of the BS, the following discrete constraint needs to be considered,
\begin{align}\label{redd}
\mathbf{s}_b^T
\mathbf{U}\mathbf{g}_b\geq 0,~\forall b\in\mathcal{B}.
\end{align}

\subsubsection{Minimum-Distance Constraint Between Discrete Positions}
To avoid overlap and mutual coupling between 6DMA surfaces at two different discrete positions, a minimum distance $d_{\min}$ is required.
We define the matrix
$\mathbf{D}\in\mathbb{C}^{M \times M}$, whose element $[\mathbf{D}]_{m,m'}=\|\mathbf{q}_m-\mathbf{q}_{m'}\|_2$ in the $m$-th row and $m'$-th column denotes the distance between the $m$-th discrete position and the $m'$-th discrete position (assuming $[\mathbf{D}]_{m, m}=0, \forall m$ for simplicity). Then, the minimum
distance constraint between any pair of 6DMA surfaces can be formulated as
\begin{align}\label{red}
\mathbf{s}_b^T
\mathbf{D}\mathbf{s}_v\geq d_{\min},~b\neq v,\forall b,v\in\mathcal{B}.
\end{align}
\subsection{Channel Model}
We consider uplink multiuser transmission, where a random number (denoted by $K$) of users (each equipped with a single FPA) are spatially distributed in a given cell and the users transmit independent messages to the 6DMA-BS of the cell.

A general multipath channel from each user to the BS is assumed.
The position vectors and rotation vectors selected for all $B$ 6DMA surfaces can be expressed as
\begin{align}
{\mathbf{q}}&=[\mathbf{q}_{i_1}^T,\mathbf{q}_{i_2}^T,\cdots,\mathbf{q}_{i_B}
^T]^T\in \mathbb{R}^{3B\times 1},\label{lk}\\
\mathbf{u}&=[\mathbf{u}_{j_1}^T,\mathbf{u}_{j_2}^T,\cdots,\mathbf{u}_{j_B}^T]^T\in \mathbb{R}^{3B\times 1}. \label{lk1}
\end{align}
Thus, the multiple-access channel matrix from the $K$ users to all 6DMA surfaces at the BS is denoted by
\begin{align}\label{hji}
\!\!\!\!\!\mathbf{H}(\mathbf{q},\mathbf{u})\!=\![\mathbf{h}_1(\mathbf{q},\mathbf{u}),\mathbf{h}_2(\mathbf{q},\mathbf{u}),\cdots,
\mathbf{h}_{K}(\mathbf{q},\mathbf{u})]\!\in \!\mathbb{C}^{NB\times K},\!\!\!\!\!
\end{align}
where $\mathbf{h}_k(\mathbf{q},\mathbf{u})\in \mathbb{C}^{NB\times 1}$ represents the channel from user $k$ to all the antennas of the $B$ 6DMA surfaces at the 6DMA-BS. $\mathbf{h}_k(\mathbf{q},\mathbf{u})$ can be expressed as
\begin{align}\label{uk}
\mathbf{h}_k(\mathbf{q},\mathbf{u})=\mathbf{G}_{k}^H(\mathbf{q},\mathbf{u})
\boldsymbol{\eta}_k,
\end{align}
with
\begin{align}
\boldsymbol{\eta}_k&=[\eta_{1,k},\eta_{2,k},\cdots,\eta_{\Gamma_{k},k}]^T\in \mathbb{C}^{\Gamma_{k}\times 1},\\
\mathbf{G}_{k}(\mathbf{q},\mathbf{u})&\!=\!
\left[\tilde{\mathbf{G}}_{k}(\mathbf{q}_{i_{1}},\mathbf{u}_{j_{1}})^H,\cdots,
\tilde{\mathbf{G}}_{k}(\mathbf{q}_{i_{B}},\mathbf{u}_{j_{B}})^H\right]
\in\mathbb{C}^{\Gamma_{k}\times NB},
\end{align}
where $\Gamma_{k}$ denotes the total number of channel paths from user $k, k\in\mathcal{K}=\{1,2,...,K\}$, to the BS, ${\eta}_{\iota, k}$ represents the channel coefficient from
user $k$ to the reference point (the center of CPU) of the BS over path $\iota$, and
\begin{align}\label{trb}
\tilde{\mathbf{G}}_{k}(\mathbf{q}_{i_{b}},\mathbf{u}_{j_{b}})&=
\left[\sqrt{g_{1,k}(\mathbf{u}_{j_{b}})}\mathbf{a}_{1,k}
(\mathbf{q}_{i_{b}},\mathbf{u}_{j_{b}}),\cdots,\right.\nonumber\\
&\left.\sqrt{g_{\Gamma_{k},k}(\mathbf{u}_{j_{b}})}\mathbf{a}_{\Gamma_{k},k}
(\mathbf{q}_{i_{b}},\mathbf{u}_{j_{b}})\right]\in\mathbb{C}^{N\times \Gamma_{k}},
\end{align}
denotes the channel matrix from the user $k$ to the $b$-th 6DMA surface, with $\mathbf{a}_{\iota,k}
(\mathbf{q}_{i_{b}},\mathbf{u}_{j_{b}})$ and $g_{\iota,k}(\mathbf{u}_{j_{b}})$ being the 6D steering vector and effective antenna gain, respectively, which are modeled in the following.
\subsubsection{6D Steering Vector}
Denote by $\phi_{\iota,k}\in[-\pi,\pi]$ and $\theta_{\iota,k}\in[-\pi/2,\pi/2]$ the azimuth and elevation angles, respectively, for the
$\iota$-th channel path between user $k$ and the BS. Accordingly, the pointing vector corresponding to direction $(\theta_{\iota,k}, \phi_{\iota,k})$ is defined as
\begin{align}\label{KM}
\!\!\!\mathbf{f}_{\iota,k}\!=\![\cos(\theta_{\iota,k})\cos(\phi_{\iota,k}), \cos(\theta_{\iota,k})\sin(\phi_{\iota,k}), \sin(\theta_{\iota,k})]^T.\!
\end{align}

By combining \eqref{nwq} and \eqref{KM}, the 6D steering vector of the $b$-th 6DMA surface for receiving a signal from user $k$ over path $\iota$ can be expressed as
\begin{align}\label{gen}
\!\!\mathbf{a}_{\iota,k}(\mathbf{q}_{i_{b}},\mathbf{u}_{j_{b}})\!=\!& \!\left[\!e^{-j\frac{2\pi}{\lambda}
\mathbf{f}_{\iota,k}^T\mathbf{r}_{i_b,1}(\!\mathbf{q}_{i_{b}},
\mathbf{u}_{j_{b}}\!)},
\!\cdots,\! e^{-j\frac{2\pi}{\lambda}\mathbf{f}_{\iota,k}^T
\mathbf{r}_{i_b,N}(\!\mathbf{q}_{i_{b}},\mathbf{u}_{j_{b}}\!)}\!\right]^T,\!\nonumber\\
&~~~~~~~~~~~~~~~~~~~~b\in\mathcal{B}, l\in\Gamma_{k}, k\in\mathcal{K},
\end{align}
which is a function of the 6DMA surface's position $\mathbf{q}_{i_{b}}$ and rotation $\mathbf{u}_{j_{b}}$. In \eqref{gen}, $\lambda$ denotes the carrier wavelength.

\subsubsection{Effective Antenna Gain}
To derive the effective antenna gain for each 6DMA surface, we project $-\mathbf{f}_{\iota,k}$ in \eqref{KM} onto the $b$-th 6DMA surface in its local CCS to obtain
\begin{align}\label{df}
\!\![\tilde{x}_{b,\iota,k},\tilde{y}_{b,\iota,k},\tilde{z}_{b,\iota,k}]^T
&\!=\!-\mathbf{R}(\mathbf{u}_{j_{b}})^T\mathbf{f}_{\iota,k}.
\end{align}
Based on \eqref{df}, we can then derive the elevation and azimuth angles of the signal direction w.r.t. the center of the $b$-th 6DMA surface in its local CCS, which are respectively given by
\begin{align}
\tilde{\theta}_{b,\iota,k}&=\pi/2-\arccos(\tilde{z}_{b,\iota,k}),\label{pp}\\
\tilde{\phi}_{b,\iota,k}&=
\arccos\left(\frac{\tilde{x}_{b,\iota,k}}{\sqrt{\tilde{x}_{b,\iota,k}^2+\tilde{y}_{b,\iota,k}^2}}
\right)\times\chi(\tilde{y}_{b,\iota,k}), \label{cc}
\end{align}
with
\begin{align}\label{ee}
\chi(\tilde{y}_{b,\iota,k})=\left\{\begin{matrix}
 1,~\tilde{y}_{b,\iota,k}\geq 0,\\
 -1, ~\tilde{y}_{b,\iota,k}< 0.
\end{matrix}\right.
\end{align}

Then, the effective antenna gain for the $b$-th 6DMA surface in the direction $(\tilde{\theta}_{b,\iota,k}, \tilde{\phi}_{b,\iota,k})$
in the linear scale is defined as
\begin{align}\label{gm}
g_{\iota,k}(\mathbf{u}_{j_{b}})=10^{\frac{A(\tilde{\theta}_{b,\iota,k}, \tilde{\phi}_{b,\iota,k})}{10}},~b\in\mathcal{B}, l\in\Gamma_{k}, k\in\mathcal{K},
\end{align}
where $A(\tilde{\theta}_{b,\iota,k}, \tilde{\phi}_{b,\iota,k})$  represents the corresponding effective antenna gain in the scale of dBi, which depends on the radiation pattern of the adopted antenna (details are provided in Section VI).

\section{Problem Formulation}
We denote the multiple-access channels from $K$ users to the 6DMA surface for all possible discrete positions and rotations as
\begin{align}\label{ff}
\bar{\mathbf{H}}=[\bar{\mathbf{H}}_{1,1}^T, \cdots, \bar{\mathbf{H}}_{1,L}^T, \cdots, \bar{\mathbf{H}}_{m,l}^T, \cdots, &\bar{\mathbf{H}}_{M,1}^T, \cdots, \bar{\mathbf{H}}_{M,L}^T]^T\nonumber\\
&\in \mathbb{C}^{NML\times K},
\end{align}
where
$\bar{\mathbf{H}}_{m,l}\in\mathbb{C}^{N\times K}, m\in \mathcal{M}, l\in \mathcal{L}$, is the channel matrix from the $K$ users to the 6DMA surface located at the $m$-th discrete position and having the $l$-th discrete rotation, which can be obtained from \eqref{hji} by replacing $\mathbf{q}$ and $\mathbf{u}$ with $\mathbf{q}_m$ and $\mathbf{u}_l^{(m)}$, respectively.
It can be shown that $\mathbf{H}(\mathbf{q},\mathbf{u})$ in \eqref{hji} can also be represented in terms of $\bar{\mathbf{H}}$ as
\begin{align}\label{hh}
 \mathbf{H}(\mathbf{q},\mathbf{u})=
(\mathbf{Q}\otimes\mathbf{I}_{N})\bar{\mathbf{H}},
\end{align}
where $\mathbf{Q}$ is a $B \times ML$ matrix whose rows are defined as
\begin{align}\label{ghy}
[\mathbf{Q}]_{b,:}= \text{vec}(\mathbf{g}_b\mathbf{s}_b^T).
\end{align}

Based on \eqref{hh}, the received signals at the BS are given by
\begin{align}\label{ly}
\mathbf{y}&=\mathbf{H}(\mathbf{q},\mathbf{u})
\mathbf{x}+\boldsymbol{\varpi}\\
&=(\mathbf{Q}\otimes\mathbf{I}_{N})\bar{\mathbf{H}}\mathbf{x}+\boldsymbol{\varpi},
\end{align}
where $\mathbf{x}=\sqrt{p}[x_1, x_2,\cdots, x_{K}]^T\in \mathbb{C}^{K\times 1}$, with $x_k$ representing the transmit signal of user $k$ having the average power normalized to one; $p$ is the transmit power of a user, which is assumed to be identical for all users; and $\boldsymbol{\varpi}\sim\mathcal{CN}(\mathbf{0}_{NB},\sigma^2\mathbf{I}_{NB})$ denotes the complex additive white Gaussian noise (AWGN) vector at the BS with zero mean and average power $\sigma^2$.

Let $\mathbf{z}=[\mathbf{s}_1^T,\mathbf{s}_2^T,\cdots,\mathbf{s}_B^T,
\mathbf{g}_1^T,\mathbf{g}_2^T,\cdots,\mathbf{g}_B^T]^T$. Since the number of users $K$ and their locations in the cell are both random, we consider the average network capacity as a metric to evaluate the communication performance of the 6DMA system, which is given by
\begin{align}\label{pc0}
C_{\rm avg}(\mathbf{z})=\mathbb{E}\left[{C}(\mathbf{z})\right],
\end{align}
with
\begin{align}
&\!\!\!C(\mathbf{z})\!=\!\log_2 \det \!\left(\!\mathbf{I}_{NB}\!+\!\frac{p}{\sigma^2}(\mathbf{Q}\otimes\mathbf{I}_{N})\bar{\mathbf{H}}
\bar{\mathbf{H}}^H(\mathbf{Q}\otimes\mathbf{I}_{N})^H\!\right)\!\\
&=\log_2 \det
\left(\mathbf{I}_{K}+\frac{p}{\sigma^2}[\bar{\mathbf{H}}^H(\mathbf{Q}^T\mathbf{Q}
\otimes\mathbf{I}_{N})\bar{\mathbf{H}}\right) \label{new1}\\
&=\log_2 \det
\!\left(\!\mathbf{I}_{K}+\frac{p}{\sigma^2}[\bar{\mathbf{H}}^H
\!\left(\!\mathrm{diag}\!\left(\!{\rm{vec}}\!\left(\!\sum_{b=1}^B\mathbf{s}_b\mathbf{g}_b^T
\!\right)\!\right)\!
\otimes\mathbf{I}_{N}\!\right)\!\bar{\mathbf{H}}\!\right)\!.\label{ne3}
\end{align}
In \eqref{pc0}, the expectation is taken w.r.t. the random channel matrix $\bar{\mathbf{H}}$ (due to the random number of users $K$ and their random locations),
the equality \eqref{new1} holds since $\det(\mathbf{I}_{p}+\mathbf{B}\mathbf{C})=\det(\mathbf{I}_{q}+\mathbf{C}\mathbf{B})$ for $\mathbf{B}\in \mathbb{C}^{p\times q}$ and $\mathbf{C}\in \mathbb{C}^{q\times p}$, and \eqref{ne3} is derived by using the following property of $\mathbf{Q}$ based on \eqref{ghy}:
\begin{align}
    \mathbf{Q}^T\mathbf{Q}&=\mathrm{diag}\left(
{\rm{vec}}\left(\sum_{b=1}^B\mathbf{g}_b\mathbf{s}_b^T\right)
\right).\label{3e1}
\end{align}

Next, we aim to find the optimal position indicator
vectors and rotation indicator vectors for all $B$ 6DMA surfaces in $\mathbf{z}$ to maximize the average network capacity given in \eqref{pc0}, subject to their practical discrete rotation/position constraints given in Section II-A. Accordingly, the optimization problem is formulated as
\begin{subequations}
\label{MG3}
\begin{align}
\text{(P1)}~&~\mathop{\max}\limits_{\mathbf{z}}~~C_{\rm avg}(\mathbf{z})
\\
\text {s.t.}
~&~(\mathbf{s}_v^T-\mathbf{s}_b^T)
\mathbf{U}\mathbf{g}_b\leq 0, b\neq v,\forall b,v\in\mathcal{B},\label{zh1}\\
~&~\mathbf{s}_b^T
\mathbf{U}\mathbf{g}_b\geq 0,~\forall b\in\mathcal{B},\label{zh}\\
~&~\mathbf{s}_b^T
\mathbf{D}\mathbf{s}_v\geq d_{\min},~b\neq v,\forall b,v\in\mathcal{B},\label{M20}\\
~&~\mathbf{1}^T\mathbf{s}_b =1, ~\forall b\in\mathcal{B}, \label{M1}\\
~&~\mathbf{1}^T\mathbf{g}_b =1, ~\forall b\in\mathcal{B}, \label{M10}\\
~&~ [\mathbf{s}_b]_m\in\{0,1\},~\forall m\in\mathcal{M}, \forall b\in\mathcal{B}, \label{M2}\\
~&~ [\mathbf{g}_b]_l\in\{0,1\},~\forall l\in\mathcal{L}, \forall b\in\mathcal{B}. \label{M8}
\end{align}
\end{subequations}
As previously discussed, constraint \eqref{zh1} avoids mutual signal
reflections between different 6DMA surfaces, constraint \eqref{zh} prevents signal blockage by
the CPU of the BS, while \eqref{M20} avoids overlapping and mutual coupling between different 6DMA surfaces. Constraints \eqref{M2} and \eqref{M8} enforce that the elements of the position indicator vectors and rotation indicator vectors are either 1 or 0 depending on whether the corresponding position/rotation is selected by the $b$-th 6DMA surface or not. Meanwhile, constraints \eqref{M1} and \eqref{M2} together
ensure that each position is not selected by more than one 6DMA surface, and constraints \eqref{M10} and \eqref{M8} together ensure that each 6DMA surface can only select a single rotation at any discrete position.

It is important to note that the objective function of problem (P1) requires the statistical knowledge of the multiple-access channel matrix $\bar{\mathbf{H}}$ in \eqref{ne3}, which may or may not be available in practice. Thus, we propose two approaches to solve (P1) under different assumptions on the statistical channel knowledge. Specifically, for the case that statistical channel knowledge is available, we propose an offline optimization solution for (P1) by approximating the average network capacity via Monte
Carlo simulation. In contrast, for the case without statistical channel knowledge beforehand, we propose an alternative online optimization approach for solving (P1), based on the achievable sum-rate values measured for different combinations of discrete positions/rotations of the 6DMA surfaces. This circumvents the need for any {\it a priori} statistical channel knowledge or explicitly estimating it in real time. In the following two sections, we discuss the above solution approaches in detail, respectively.

\section{Offline Optimization with Statistical Channel Knowledge}
In this section, we propose our offline optimization algorithm to solve (P1) for maximizing the average network capacity based on the statistical channel knowledge which is assumed to be known. In particular, we first apply the Monte Carlo method to obtain an approximation of $C_{\rm avg}(\mathbf{z})$ given
in \eqref{pc0}, as its exact expression cannot be derived analytically. This involves generating $\Omega$ independent realizations of the number of users, $K$, and their locations, and then averaging the corresponding achievable sum-rates over all realizations. Thus, the average network capacity in \eqref{pc0} can be approximated as
\begin{align}
\tilde{C}(\mathbf{z})=\frac{1}{
\Omega}\sum_{\tau=1}^{
\Omega}
{C}_\tau(\mathbf{z}). \label{pcb}
\end{align}
where ${C}_{\tau}(\mathbf{z})={C}(\mathbf{z})|
\bar{\mathbf{H}}_{\tau}$ denotes the achievable sum-rate for the $\tau$-th realization given the channel matrix $\bar{\mathbf{H}}_{\tau}$.
Then, problem (P1) reduces to
\begin{subequations}
\label{MG32}
\begin{align}
\text{(P2)}~&~\mathop{\max}\limits_{\mathbf{z}}~~\tilde{C}(\mathbf{z})
\\
\text {s.t.}~&~\eqref{zh1}, \eqref{zh}, \eqref{M20}, \eqref{M1},
\eqref{M10}, \eqref{M2},\eqref{M8}.
\end{align}
\end{subequations}
We observe that the objective function/constraints \eqref{zh1}--\eqref{M20} of (P2) are non-convave/non-convex w.r.t. $\mathbf{z}$ (i.e., $\mathbf{s}_b$ and $\mathbf{g}_b, b\in\mathcal{B}$), and the variables in $\mathbf{z}$ are binary. Thus, (P2) is a non-convex integer programming problem that is difficult to be optimally solved efficiently, when $M$ (or $L$) and $B$ are large values.

{\bf{Remark 1}}: It is important to note that (P2) differs significantly from traditional antenna selection optimization \cite{as9,as10,sele}. In antenna selection optimization, the active antennas are selected from a given number of antennas at fixed positions. However, in (P2), not only the positions but also the rotations need to be selected for multiple 6DMA surfaces, which are also subject to practical position/rotation constraints, thus making the problem much more complicated to solve.

To solve problem (P2), we first reformulate the non-convex quadratic inequality constraints \eqref{zh1}--\eqref{M20} into convex linear inequality constraints by exploiting the following lemma, the proof of which is given in \cite{dis0} and thus omitted here.

{\bf{Lemma 1}}: Inequality constraint \eqref{M20} is equivalent to the following linear inequality constraints,
\begin{align}
&\sum_{m\in\mathcal{M}}\sum_{l\in\mathcal{M}}[\mathbf{D}]_{m,l}\bar{w}_{b,v,m,l}\geq d_{\min},
~b\neq v, \forall b,v\in\mathcal{B}, \label{h5}\\
&\bar{w}_{b,v,m,l}\leq \min \{[\mathbf{s}_b]_m,[\mathbf{s}_v]_l\},~b\neq v, \forall b,v\in\mathcal{B}, \forall m,l\in\mathcal{M}, \label{h6}\\
&\bar{w}_{b,v,m,l}\geq [\mathbf{s}_b]_m+[\mathbf{s}_v]_l-1, ~b\neq v, \forall b,v\in\mathcal{B}, \forall m,l\in\mathcal{M}, \label{h7}
\end{align}
where $\bar{w}_{b,v,m,l}$ is a binary auxiliary variable. For the sake of notational simplicity, we define a binary vector $\bar{\mathbf{w}}=[\bar{w}_{1,2,1,1},\cdots,\bar{w}_{b,v,m,l},\cdots,\bar{w}_{B-1,B,M,M}],~b\neq v, \forall b,v\in\mathcal{B}, \forall m,l\in\mathcal{M}$ to collect all binary auxiliary variables.

Similarly, the inequality constraint \eqref{zh1} is equivalent to the following linear inequality constraints,
\begin{align}
&\sum_{m\in\mathcal{M}}\sum_{l\in\mathcal{L}}[\mathbf{U}]_{m,l}w_{b,v,m,l}-
\sum_{m\in\mathcal{M}}\sum_{l\in\mathcal{L}}[\mathbf{U}]_{m,l}w_{b,b,m,l}\leq 0,\nonumber\\
&~~~~~~~~~~~~~~~~~~~~~~~~~~~~~~~~~~~~~~~~~~~b\neq v, \forall b,v\in\mathcal{B}, \label{bh5}\\
&w_{b,v,m,l}\leq \min \{[\mathbf{s}_v]_m,[\mathbf{g}_b]_l\}, \forall b,v\in\mathcal{B},\forall m \in\mathcal{M},\forall l\in\mathcal{L}, \label{bbh6}\\
&w_{b,v,m,l}\geq [\mathbf{s}_v]_m+[\mathbf{g}_b]_l-1, \forall b,v\in\mathcal{B}, \forall m\in\mathcal{M}, \forall l\in\mathcal{L}, \label{bbh7}
\end{align}
where $w_{b,v,m,l}$ is a binary auxiliary variable and $\mathbf{w}=[w_{1,2,1,1},\cdots,w_{b,v,m,l},
\cdots,w_{B,B,M,L}], \forall b,v\in\mathcal{B}, \forall m\in\mathcal{M}, \forall l\in\mathcal{L}$.

Finally, inequality constraint \eqref{zh} is equivalent to the following linear inequality constraints,
\begin{align}
&\sum_{m\in\mathcal{M}}\sum_{l\in\mathcal{L}}[\mathbf{U}]_{m,l}w_{b,b,m,l}\geq 0,~\forall b\in\mathcal{B}, \label{bbh5}\\
&\eqref{bbh6}, \eqref{bbh7}. \nonumber
\end{align}

Based on Lemma 1, (P2) can be equivalently reformulated as
\begin{subequations}
\label{jMG3}
\begin{align}
\!\!\!\!\!\!\!\!\!\!\text{(P2-1)}~&~\mathop{\max}\limits_{\mathbf{z},\mathbf{w},\bar{\mathbf{w}}}~~
\tilde{C}(\mathbf{z})\\
\text {s.t.}~&~ \eqref{M1},\eqref{M10}, \eqref{M2},\eqref{M8},\eqref{h5}, \eqref{h6},\eqref{h7},\eqref{bh5}, \eqref{bbh6},
\nonumber\\
~&~\eqref{bbh7}, \eqref{bbh5}, \\
~&~\bar{w}_{b,v,m,l}\!\in\!\{0,1\},~b\neq v, \forall b,v\!\in\!\mathcal{B},\forall m, l\!\in\!\mathcal{M}, \!\!\!\!\!\\
~&~w_{b,v,m,l}\in\{0,1\},~\forall b,v\in\mathcal{B}, \forall m\in\mathcal{M}, \forall l\in\mathcal{L}.
\end{align}
\end{subequations}

The feasible region of problem (P2-1) is a convex set (for continuous $\mathbf{z}$, $\mathbf{w}$ and $\bar{\mathbf{w}}$), while its objective function is still non-concave w.r.t. $\mathbf{z}$, which makes it difficult to solve with binary integer variables. Therefore, we apply linear programming relaxation by relaxing all the binary integer variables to be continuous over the interval [0,1] \cite{nonl1}. Consequently, (P2-1) is relaxed as follows,
\begin{subequations}
\begin{align}
\!\!\!\!\!\!\!\!\!\!\text{(P2-2)}~&~\mathop{\max}\limits_{\mathbf{z},
\bar{\mathbf{w}},\mathbf{w}}~~
\tilde{C}(\mathbf{z}) \label{jMG5}\\
\text {s.t.}~&~ \eqref{M1},
\eqref{M10}, \eqref{h5}, \eqref{h6},\eqref{h7},\eqref{bh5}, \eqref{bbh6}, \eqref{bbh7}, \eqref{bbh5},  \label{vg2}\\
~&~ [\mathbf{s}_b]_m \in[0,1],~\forall m\in\mathcal{M}, \forall b\in\mathcal{B}, \label{Mu2}\\
~&~ [\mathbf{g}_b]_l \in[0,1],~\forall l\in\mathcal{L}, \forall b\in\mathcal{B}, \label{Mu8}\\
~&~\bar{w}_{b,v,m,l}\in[0,1], ~b\neq v, \forall b,v\!\in\!\mathcal{B},\forall m, l\!\in\!\mathcal{M}, \!\!\!\!\!\label{vg}\\
~&~w_{b,v,m,l}\in[0,1],~\forall b,v\in\mathcal{B}, \forall m\in\mathcal{M}, \forall l\in\mathcal{L}.\label{vg1}
\end{align}
\end{subequations}
(P2-2) can be efficiently solved by the conditional gradient method \cite{linear, mk} sub-optimally (see Algorithm 1 for details).

After the solution of problem (P2-2) is obtained, which is denoted as $\mathbf{z}^*=[(\mathbf{s}_1^{*})^T,\cdots,(\mathbf{s}_B^{*})^T,
(\mathbf{g}_1^{*})^T,\cdots,(\mathbf{g}_B^{*})^T]^T$, we compute $\mathbf{S}^*=\sum_{b=1}^B\mathbf{s}_b^*(\mathbf{g}_b^{*})^ {T}\in\mathbb{R}^{M\times L}$.
To obtain the indices of optimized 3D positions and 3D rotations for $B$  6DMA surfaces, we define the position utility vector and the rotation utility vector, respectively, as follows,
\begin{align}\label{uti}
&\mathbf{J}_{\mathrm{f}}=\left[\sum_{l=1}^{L}
[\mathbf{S}^*]_{1,l},\sum_{l=1}^{L}[\mathbf{S}^*]_{2,l},
\cdots,\sum_{l=1}^{L}[\mathbf{S}^*]_{M,l}\right]^T\in\mathbb{R}^{M\times 1},\\
&\bar{\mathbf{J}}_{\mathrm{f}}^{(m)}=[\mathbf{S}^*]_{m,:}\in\mathbb{R}^{L\times 1}, m\in \mathcal{M}.
\end{align}

Then, the indices of optimized 3D
positions of $B$ 6DMA surfaces are given by
\begin{align}\label{pp}
i_1^*,i_2^*,\cdots,i_B^*=\arg \mathop{\max}\limits_{\text{Top}~ B}~\mathbf{J}_{\mathrm{f}},
\end{align}
where $\arg \mathop{\max}\limits_{\text{Top}~ B}$ returns the indices of the largest $B$ values in $\mathbf{J}_{\mathrm{f}}$, in a decreasing order.

Moreover, the indices of optimized 3D
rotations of $B$ 6DMA surfaces are given by
\begin{align}\label{pp1}
j_b^*=\arg \mathop{\max}\limits_{l}~[\bar{\mathbf{J}}_{\mathrm{f}}^{(i_b^*)}]_{l},~b\in\mathcal{B}.
\end{align}

The details of the offline optimization algorithm for solving (P1) or equivalently (P2-2) are provided in Algorithm 1. Let $\bar{K} = \max(K_1, K_2, \ldots, K_{\Omega})$ with $K_{\tau}$ being the number of users in the $\tau$-th Monte Carlo channel realization. The overall complexity order of the proposed  Algorithm 1 is $\mathcal{O}(T_{\mathrm{f}} M^2N^2L^2\bar{K}\Omega)$, where $T_{\mathrm{f}}$ denotes the maximum number of iterations in Algorithm 1.
Algorithm 1 is convergent because the gradient-based search ensures that the objective value of (P2-2) does not decrease over the iterations. Furthermore, since (P2-2) is a constrained problem, its objective value is upper-bounded by a finite value.

\begin{algorithm}[t!]
\caption{Offline Optimization Algorithm for Solving (P2-2).}
\label{alg0}
\begin{algorithmic}[1]
\STATE \textbf{Input}: $B$, $N$, $M$, $L$, $\lambda$, and $T_{\mathrm{f}}$.  \\
\STATE {Initialization}: $t=0$, $\mathbf{z}^{(0)}$, ${\mathbf{w}}^{(0)}$, and $\bar{\mathbf{w}}^{(0)}$.
 \WHILE{$t<T_{\mathrm{f}}$ }

    \STATE Compute the gradients of $\tilde{C}(\mathbf{z})$ as
    \begin{align}
\!\!\![\nabla\tilde{C}(\mathbf{z}^{(t)})]_v\!=
\!&\lim_{\varepsilon\rightarrow 0}\frac{\tilde{C}(\mathbf{z}^{(t)}+\varepsilon\mathbf{e}^{v})
\!-\!\tilde{C}(\mathbf{z}^{(t)})}{\varepsilon},1 \leq v \leq BML,\nonumber
    \end{align}
where $\mathbf{e}^{v}\in \mathbb{R}^{BML}$ is a vector with a one as the $v$-th element and zeros elsewhere;

    \STATE Construct the direction-finding vectors $\mathbf{z}_{\mathrm{f}}^{(t)}$, $\bar{\mathbf{w}}_{\mathrm{f}}^{(t)}$, and $\mathbf{w}_{\mathrm{f}}^{(t)}$ by solving the following problem using linprog \cite{lin},
\begin{align}
\mathop{\min}\limits_{\mathbf{z},\bar{\mathbf{w}},\mathbf{w}}
~&~-\nabla\tilde{C}(\mathbf{z}^{(t)})^T(\mathbf{z}-\mathbf{z}^{(t)}),\nonumber\\
\text {s.t.}~&~\eqref{vg2}, \eqref{Mu2}, \eqref{Mu8}, \eqref{vg}, \eqref{vg1}.\nonumber
\end{align}
    \STATE Update the indicator vectors for the 3D positions and 3D rotations and the auxiliary variables according to:
\begin{align}
\mathbf{z}^{(t+1)}&=\mathbf{z}^{(t)}+\varsigma^{(t)}
(\mathbf{z}_{\mathrm{f}}^{(t)}-\mathbf{z}^{(t)}),\nonumber\\
\mathbf{w}^{(t+1)}&=\mathbf{w}^{(t)}+\varsigma^{(t)}
(\mathbf{w}_{\mathrm{f}}^{(t)}-\mathbf{w}^{(t)}),\nonumber\\
\bar{\mathbf{w}}^{(t+1)}&=\bar{\mathbf{w}}^{(t)}+\varsigma^{(t)}
(\bar{\mathbf{w}}_{\mathrm{f}}^{(t)}-\bar{\mathbf{w}}^{(t)}),\nonumber
\end{align}
where $\varsigma^{(t)}\in(0,1]$ is the step size;

     \STATE Update $t=t+1$;
     \ENDWHILE
 \STATE Obtain $\{\mathbf{s}_b^*\}_{b\in \mathcal{B}}$ and $\{\mathbf{g}_b^*\}_{b\in \mathcal{B}}$ from $\mathbf{z}^{(t)}$;
 \STATE Determine $\{i_b^*\}_{b\in \mathcal{B}}$ according to \eqref{pp};
 \STATE Determine $\{j_b^*\}_{b\in \mathcal{B}}$ according to \eqref{pp1};
\STATE \textbf{Output}: $\{i_b^*\}_{b\in \mathcal{B}}$ and $\{j_b^*\}_{b\in \mathcal{B}}$.
\end{algorithmic}
\end{algorithm}

\section{Online Optimization Without Statistical Channel Knowledge}
In this section, we propose an online algorithm for solving (P1)
in the case without any a priori statistical channel knowledge. In particular, we solve (P1) based on a set of achievable sum-rate values measured at the BS for different combinations of discrete positions/rotations of 6DMA surfaces\footnote{It is worth noting that the proposed online algorithm applies to other performance metrics measurable in practical systems in addition to the sum-rate, which is used in this paper just as an example for obtaining the system performance limit.}.

\subsection{Generation of Discrete Positions and Rotations}
Prior to introducing our online algorithm, we need to generate offline $M$ discrete positions, $\mathbf{q}_m, m\in\mathcal{M}$, and $L$ discrete rotations, $\mathbf{u}_{l}^{(m)}, l\in\mathcal{L}$, within the given 6DMA-BS site space, $\mathcal{C}$. For convenience, we define each generated
position-rotation pair as
\begin{align}\label{set}
\mathbf{v}_{l}^{(m)}=[\mathbf{q}_{m}^T,(\mathbf{u}_{l
}^{(m)})^T]^T\in\mathbb{C}^{6\times 1}, ~m\in \mathcal{M}, l
\in \mathcal{L}.
\end{align}
Then, the set of all generated position-rotation pairs
is given by
\begin{align}\label{bb}
\mathcal{V}=\{\mathbf{v}_{1}^{(1)},\cdots, \mathbf{v}_{L}^{(1)}, \cdots, \mathbf{v}_{l}^{(m)}, \cdots, \mathbf{v}_{1}^{(M)}, \cdots,\mathbf{v}_{L}^{(M)}\},
\end{align}
with $|\mathcal{V}|_\mathrm{c}=ML$. In addition, $\mathcal{V}$ is generated to satisfy the following condition: any two different position-ration pairs in $\mathcal{V}$ satisfy all the practical discrete position/rotation constraints given in Section II-A.

\begin{figure}[t!]
\centering
\setlength{\abovecaptionskip}{0.cm}
\includegraphics[width=2.1in]{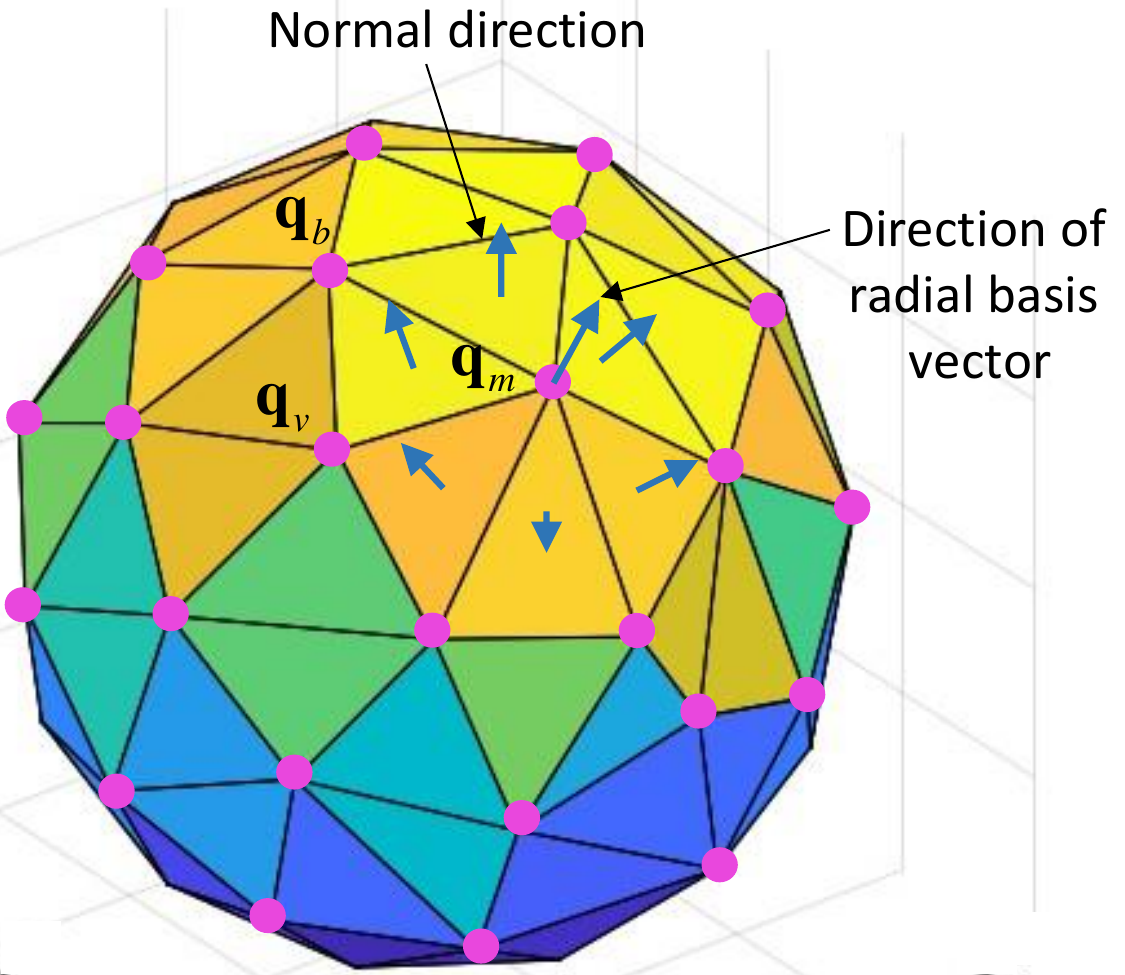}
\caption{Illustration of the discrete positions and rotations generated on a spherical surface (circles represent discrete positions, while arrows represent the possible normal vectors of 6DMA surface at position $\mathbf{q}_m$).}
\label{unknownchannel}
\end{figure}

To achieve this goal, the discrete positions of 6DMA surfaces are only taken from points on the surface of a sphere centered at the CPU and with the largest possible radius within $\mathcal{C}$. Then, we
uniformly generate $M$ discrete positions on this spherical surface by applying the Fibonacci Sphere scheme \cite{trip}, as shown in Fig. \ref{unknownchannel}, which provides a uniform distribution of positions on a sphere. By carefully selecting the value of $M$, we can ensure that the minimum distance $d_{\min}$ between any two positions is guaranteed. Next, given the $M$ discrete positions, we need to determine $L\geq 1$ discrete rotations for each discrete position. To meet the two discrete rotation constraints given in Section II-A, we construct  the convex hull of the $M$ generated positions (see Fig. \ref{unknownchannel}). The outward normal directions of all spatial surfaces on the convex hull that include position $\mathbf{q}_m$, as well as the direction of the radial basis vector of $\mathbf{q}_m, m\in\mathcal{M}$, can be used to generate desired rotations of the 6DMA surface at discrete position $\mathbf{q}_m$, with the details given in the following.

First, we consider the case where the $x'$-axis (in the local CCS) of the 6DMA surface at position $\mathbf{q}_m$ aligns with the direction of the radial basis vector of $\mathbf{q}_m$ (see Fig. \ref{unknownchannel}). For convenience, we convert the global CCS of $\mathbf{q}_m$ to the equivalent spherical coordinates $(r_m,\omega_m,\zeta_m)$, where $r_m$, $\omega_m$, and $\zeta_m$ represent the radius, polar angle, and azimuthal angle, respectively.
Then, we orient the $x'$-axis along the direction of the radial basis vector $\mathbf{a}_{r_m}=[s_{\omega_m} c_{\zeta_m}, s_{\omega_m} s_{\zeta_m}, c_{\omega_m}]^T$ in spherical coordinate. Similarly, we align the $y'$-axis along the direction of the azimuthal basis vector $\mathbf{a}_{\zeta_m}=[-s_{\omega_m} s_{\zeta_m}, s_{\omega_m} c_{\zeta_m}, 0]^T$, and the $z'$-axis along the direction of the polar basis vector $\mathbf{a}_{\omega_m}=[c_{\omega_m} c_{\zeta_m},c_{\omega_m} s_{\zeta_m},-s_{\omega_m}]^T$. Next, by substituting $\mathbf{q}_m$ and the corresponding basis vectors into \eqref{nwq}, the rotation matrix of the 6DMA surface at the $m$-th discrete position with the $l$-th discrete rotation can be determined as
\begin{align}\label{gt}
\mathbf{R}(\mathbf{u}_{m}^{(l)}) = \left[\mathbf{a}_{r_m}, \mathbf{a}_{\zeta_m}, \mathbf{a}_{\omega_m}\right].
\end{align}

Second, we consider the case where the $x'$-axis aligns with the outward normal vector of a spatial surface (see Fig. \ref{unknownchannel}). Specifically, for each spatial surface including position $\mathbf{q}_m$, we denote two other positions on this surface as $\mathbf{q}_{b}$ and $\mathbf{q}_v$, $m, b, v\in{\mathcal{M}}$. We then determine two vectors passing through these three points, denoted by $\boldsymbol{\vartheta}_m = \mathbf{q}_m - \mathbf{q}_{b}$ and $\bar{\boldsymbol{\vartheta}}_m = \mathbf{q}_v-\mathbf{q}_m$. The cross product of these two vectors yields the normal vector of this surface, expressed as
$\tilde{\mathbf{n}}_m=\frac{\boldsymbol{\vartheta}_m\wedge\bar{\boldsymbol{\vartheta}}_m}{\|\boldsymbol{\vartheta}_m\wedge\bar{\boldsymbol{\vartheta}}_m\|_2}$.
Subsequently, we orient the $x'$-axis along the direction of $\tilde{\mathbf{n}}_m$, align the $y'$-axis along the direction of $\frac{\mathbf{q}_v-\mathbf{q}_{b}}{\|\mathbf{q}_v-\mathbf{q}_{b}\|_2}$, and align $z'$-axis along the direction of $\tilde{\mathbf{n}}_m \wedge\frac{\mathbf{q}_v-\mathbf{q}_{b}}{\|\mathbf{q}_v-\mathbf{q}_{b}\|_2}$.
Next, by substituting $\mathbf{q}_m$ and the aforementioned three directions into \eqref{nwq}, the rotation matrix of the 6DMA surface at the $m$-th discrete position with the $l$-th discrete rotation can be determined as
\begin{align}\label{jio}
\mathbf{R}(\mathbf{u}_{l}^{(m)}) = \left[\tilde{\mathbf{n}}_m, \frac{\mathbf{q}_v-\mathbf{q}_{b}}{\|\mathbf{q}_v-\mathbf{q}_{b}\|_2},\tilde{\mathbf{n}}_m \wedge\frac{\mathbf{q}_v-\mathbf{q}_{b}}{\|\mathbf{q}_v-\mathbf{q}_{b}\|_2}\right].
\end{align}
Based on $\mathbf{R}(\mathbf{u}_{l}^{(m)})$ in \eqref{gt} and \eqref{jio}, the $l$-th rotation $\mathbf{u}_{l}^{(m)}, l\in\mathcal{L}$, corresponding to discrete position $\mathbf{q}_m$ can be obtained as \cite{rot3}:
\begin{align}\label{hrhhre}
\mathbf{u}_{l}^{(m)} = \left[\begin{matrix}
\arctan2\left([\mathbf{R}(\mathbf{u}_{l}^{(m)})]_{2,3},[\mathbf{R}(\mathbf{u}_{l}^{(m)})]_{3,3}\right)\\ -\arcsin\left([\mathbf{R}(\mathbf{u}_{l}^{(m)})]_{1,3}\right)\\ \arctan2\left([\mathbf{R}(\mathbf{u}_{l}^{(m)})]_{1,2},[\mathbf{R}(\mathbf{u}_{l}^{(m)})]_{1,1}\right)
\end{matrix}\right].
\end{align}

\subsection{CSM-Based Online Algorithm}
In this subsection, we propose a CSM-based online algorithm to solve (P1) for finding the optimal 3D discrete positions and 3D discrete rotations of $B$ 6DMA surfaces.

First, we randomly and independently generate $T$ sample sets each consisting  of $B$ discrete position-rotation pairs, with their indices given by $[i_{b,t}, j_{b,t}]^T, b\in\mathcal{B}, t\in \mathcal{T}$ with $\mathcal{T}=\{1,2,\cdots,T\}$. Specifically, for each sample set, we begin by selecting $B$ positions from the $M$ discrete positions (offline generated in Section V-A) in $\mathcal{M}$ with equal probability. Then, for each selected position, we choose one rotation from the $L$ discrete rotations offline generated in $\mathcal{L}$ with equal probability. Thus, each sample set is given by
\begin{align}\label{ran}
{\mathcal{V}}_t=\left\{\mathbf{v}_{j_{1,t}}^{(i_{1,t})},\mathbf{v}_{j_{2,t}}^{(i_{2,t})},\cdots, \mathbf{v}_{j_{B,t}}^{(i_{B,t})}\right\}\in \mathcal{V}, ~t\in\mathcal{T},
\end{align}
with $|\mathcal{V}_t|_{\mathrm{c}}=B$.

\begin{algorithm}[t!]
\caption{CSM-based Online Optimization Algorithm for Solving (P1).}
\label{alg3}
\begin{algorithmic}[1]
\STATE \textbf{Input}: $\mathcal{V}$, $B$, $T$.

    \FOR {$t = 1$ to $T$}
        \STATE Randomly generate ${\mathcal{V}}_t$ from ${\mathcal{V}}$ given in \eqref{bb};
      \STATE Measure the corresponding sum-rate value  $C({q},\mathbf{u})^{t}$ with ${\mathcal{V}}_t$;
         \ENDFOR
    \FOR {$m=1,\cdots, M$}
  \FOR {$l=1,\cdots, L$}
    \STATE Compute $\widehat{\mathbb{E}}\left[C(\mathbf{q},\mathbf{u})^{(t)}
\big|t\in \mathcal{G}_{m,l}\right]$ according to \eqref{ep};
    \ENDFOR
\ENDFOR
 \STATE Determine $\{i_b^*\}_{b\in \mathcal{B}}$ according to \eqref{ppo};
 \STATE Determine $\{j_b^*\}_{b\in \mathcal{B}}$ according to \eqref{pp1o};
\STATE \textbf{Output}: $\{i_b^*\}_{b\in \mathcal{B}}$ and $\{j_b^*\}_{b\in \mathcal{B}}$.
\end{algorithmic}
\end{algorithm}

Let $\mathcal{G}_{m,l}\subseteq \mathcal{T}$ be a subset of indices under which the sample set ${\mathcal{V}}_t$ includes the position-rotation pair $\mathbf{v}_{l}^{(m)}$, i.e.,
\begin{align}\label{low}
\mathcal{G}_{m,l}=\{t \in\mathcal{T}~\text{such that}~ \mathbf{v}_{l}^{(m)}\in {\mathcal{V}}_t\}.
\end{align}

We then compute the sample mean of the sum-rate conditioned on each $\mathcal{G}_{m,l}$ as\footnote{If $|\mathcal{G}_{m,l}|_{\mathrm{c}}=0$, then we set the conditional sample mean value to zero without loss of generality.}
\begin{align}\label{ep}
\!\!\!\!\widehat{\mathbb{E}}\left[C(\mathbf{q},\mathbf{u})^{(t)}
\big|t\in \mathcal{G}_{m,l}\right]=\frac{1}{|\mathcal{G}_{m,l}|_{\mathrm{c}}}
\sum_{t\in \mathcal{G}_{m,l}}C(\mathbf{q},\mathbf{u})^{(t)},
\end{align}
where
\begin{align}\label{qu}
C(\mathbf{q},\mathbf{u})^{(t)}=\log_2 \det \left(\mathbf{I}_{NB}+\frac{p}{\sigma^2}
\mathbf{H}(\mathbf{q},\mathbf{u})\mathbf{H}(\mathbf{q},\mathbf{u})^H\right)
\big|
\mathcal{V}_t,
\end{align}
is the achievable sum-rate value measured with the $B$ 6DMA surfaces taking the position-rotation pairs in ${\mathcal{V}}_t$.

Intuitively, the conditional sample mean in \eqref{ep}
characterizes the average performance by setting $\mathbf{v}_{l}^{(m)}\in {\mathcal{V}}_t$. Thus, we define the rotation utility and position utility for online optimization, respectively, as follows:
\begin{align}\label{utio}
&\varphi^{(m)}=\arg \mathop{\max}\limits_{l}~ \widehat{\mathbb{E}}\left[C(\mathbf{q},\mathbf{u})^{(t)}
\big|t\in \mathcal{G}_{m,l}\right],~l\in\mathcal{L},\nonumber\\
&{\mathbf{J}}_{\mathrm{e}}=\left[\widehat{\mathbb{E}}\left[C(\mathbf{q},\mathbf{u})^{(t)}
\big|t\in \mathcal{G}_{1,\varphi^{(1)}}\right],\cdots,\right.\nonumber\\
&~~~~~~~\left.\widehat{\mathbb{E}}\left[C(\mathbf{q},\mathbf{u})^{(t)}
\big|t\in \mathcal{G}_{M,\varphi^{(M)}}\right]\right]^T.
\end{align}

Then, the indices of the optimal 3D
positions of the 6DMA surfaces are given by
\begin{align}\label{ppo}
i_1^*,i_2^*,\cdots,i_B^*=\arg \mathop{\max}\limits_{\text{Top}~ B}~{\mathbf{J}}_{\mathrm{e}}.
\end{align}
In addition, the indices of the optimal 3D
rotations of the 6DMA surfaces are given by
\begin{align}\label{pp1o}
j_b^*=\varphi^{(i_b^*)},~b\in\mathcal{B}.
\end{align}

The above CSM-based online algorithm is summarized in Algorithm 2.
The complexity of this algorithm is mainly due to the computation of the CSM values for the sum-rate, which is thus in the order of
$\mathcal{O}(TB)$.

\section{Simulation Results}
\begin{figure}[t!]
\centering
\setlength{\abovecaptionskip}{0.cm}
\includegraphics[width=3.0in]{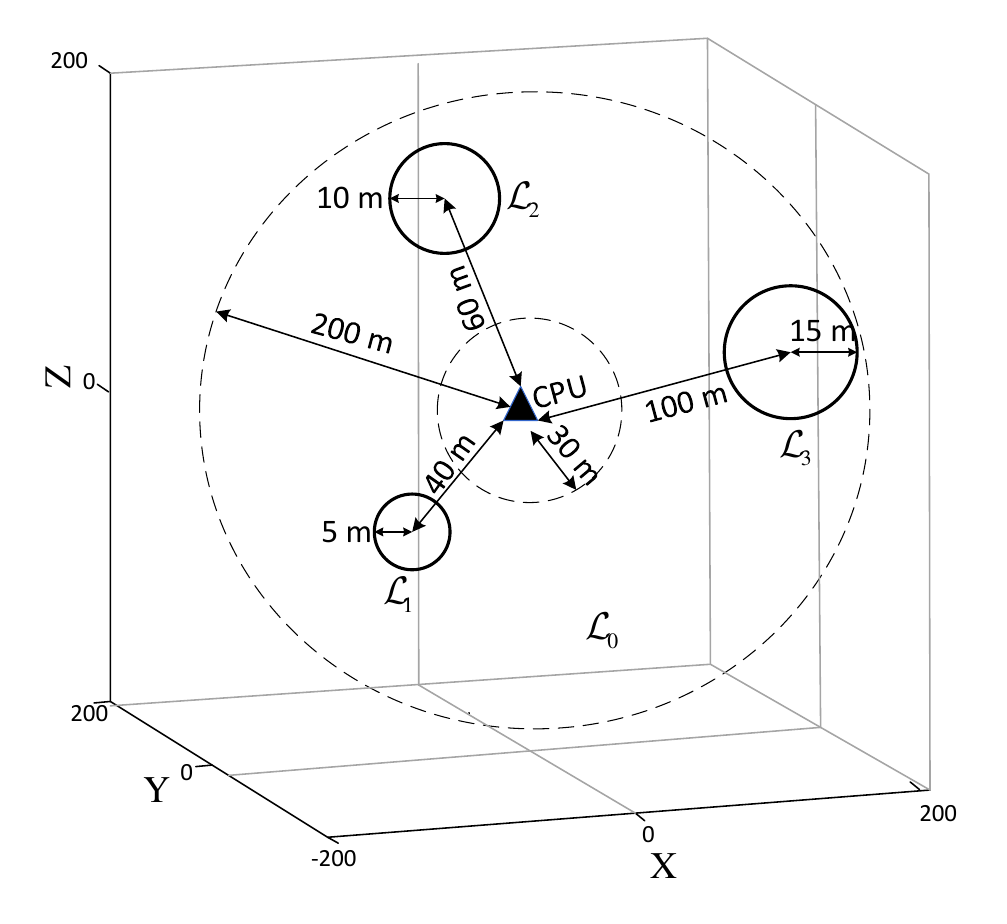}
\caption{Simulation setup.}
\label{step}
\end{figure}
In this section, numerical results are provided to validate
the performance of our proposed 6DMA-BS design with discrete position and rotation optimization. For simplicity, we adopt the line-of-sight (LoS) channel model from all users to the BS (i.e., $\Gamma_{k}=1, \forall k$) in our simulations. Unless specified otherwise, we set $N=4$, i.e., each 6DMA surface is equipped with a $2\times 2$ UPA with antenna element spacing $\frac{\lambda}{2}$, and the number of 6DMA surfaces is $B=16$.
The carrier frequency is 2.4 GHz and the wavelength is $\lambda=0.125$ m, the transmit power of each user is
$p=60$ mW, and the average number of users is $\mu=40$.
We assume that the minimum distance required
between any two discrete positions is $d_{\min}=\frac{\sqrt{2}}{2}\lambda+\frac{\lambda}{2}$, where $\frac{\sqrt{2}}{2}\lambda$ is the diagonal length of the (square) uniform UPA. For the offline optimization algorithm with statistical channel knowledge, we set the number of Monte Carlo realizations as $\Omega=100$, while for the CSM-based online algorithm without statistical channel knowledge, we set the sample size as $T=M^2L^2$.
The 6DMA-BS site space, $\mathcal{C}$, is set as a cube with 1 m side length. Furthermore, we consider a scenario where the users are located in a 3D coverage region $\mathcal{L}$, which includes three non-overlapping hotspot sub-regions, denoted by $\mathcal{L}_v, v=1,2,3$, and the remaining region, denoted by $\mathcal{L}_0$, such that $\mathcal{L}=\mathcal{L}_0\cup(\cup_{v=1}^3\mathcal{L}_v)$.
As shown in Fig. \ref{step}, $\mathcal{L}$ is set as a 3D spherical annulus with radial distances extending from 30 m to 200 m from the center of the CPU.
We further set the hotspot sub-regions $\mathcal{L}_1$, $\mathcal{L}_2$, and $\mathcal{L}_3$ as 3D spheres with their respective centers positioned at distances of 40 m, 60 m, and 100 m from the CPU center with radii of 5 m, 10 m, and 15 m, respectively. For the offline optimization, multiple independent random initializations are utilized, where the initial positions and rotations of the 6DMA surfaces are randomly chosen from the set $\mathcal{V}$, respectively.

To evaluate the impact of the spatial user distribution on the optimized 6DMA positions/rotations as well as the network capacity, we adopt the general non-homogeneous Poisson Process (NHPP) model \cite{flow1} for generating the  number of users and their locations. For this model, the probability mass function (PMF) of the number of users, $K$, follows a Poisson distribution, i.e.,
\begin{align}\label{pmf}
\mathrm{Pr}[K=\tilde{K}]=\frac{\mu^{\tilde{K}}}{\tilde{K}!}e^{-\mu},~ \tilde{K}=0,1,2, \cdots.
\end{align}
with the mean given by
\begin{align}\label{muw}
\mu=\int_{\mathcal{L}}\rho(\boldsymbol{\kappa})d\boldsymbol{\kappa},
\end{align}
where $\rho(\boldsymbol{\kappa})$ (in users/m$^3$) denotes the user density function  for location $\boldsymbol{\kappa}\in\mathcal{L}$ and is defined as
\begin{align}\label{hq}
\rho(\boldsymbol{\kappa})=\left\{\begin{matrix}
&\rho_0, &~\boldsymbol{\kappa}\in\mathcal{L}_0,\\
&\rho_0+\rho_v, &~\boldsymbol{\kappa}\in\mathcal{L}_{v},  v\in\{1,2,3\}.
\end{matrix}\right.
\end{align}
Here, $\rho_{v}\geq0$ and $\rho_0\geq0$ are constant density values for hotspot users and background/regular users \cite{cluter}, respectively, which are set to ensure that the average numbers of users in the hotspot sub-regions $\mathcal{L}_1$, $\mathcal{L}_2$, and $\mathcal{L}_3$ match the ratio of 1:2:3. Moreover, the proportion of regular users is denoted as $\xi=\frac{\int_{\mathcal{L}}\rho_0d\boldsymbol{\kappa}}{\mu}$, which is set as $\xi=0.2$, unless specified otherwise. For both online and offline optimization, the number of users and their locations are randomly generated according to the above NHPP-based user spatial distribution.
\begin{figure}[t!]
\centering
\setlength{\abovecaptionskip}{0.cm}
\includegraphics[width=3.69in]{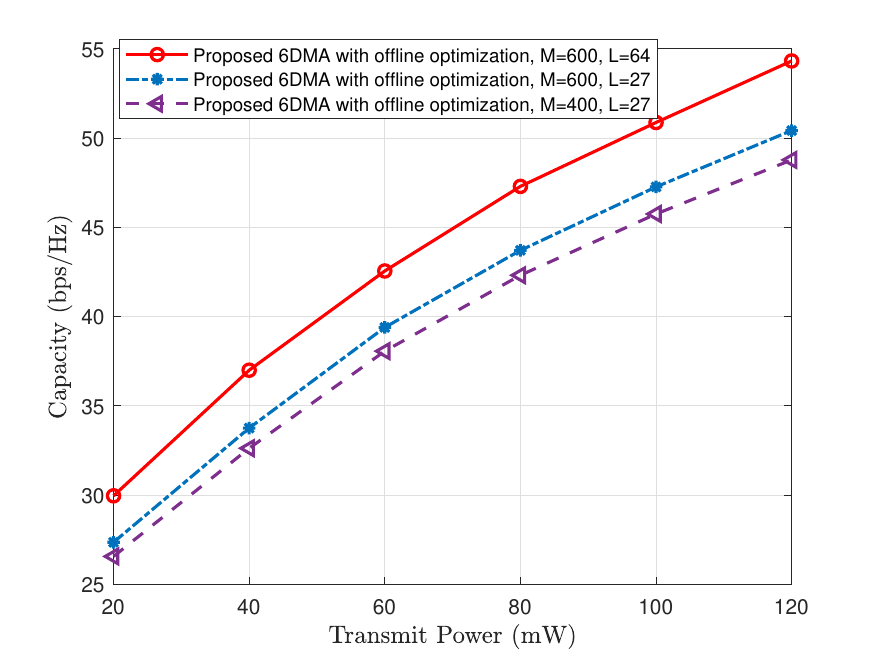}
\caption{Average network capacity versus transmit power for the proposed offline optimization algorithm.}
\label{alg_user}
\end{figure}

For online optimization, the discrete positions and rotations for the 6DMA surfaces are generated using the method given in Section V-A. In contrast, for the offline optimization, we consider a more general method to generate the discrete positions and rotations, described in the following. First, the discrete position values are obtained by uniformly quantizing the 3D BS site space $\mathcal{C}$. Then, we obtain the rotation angles $\alpha_l^{(m)}$, $\beta_l^{(m)}$, and $\gamma_l^{(m)}$, $m\in \mathcal{M}, l\in\mathcal{L}$, by uniformly sampling the corresponding interval from 0 to $2\pi$. This results in $L = Z^3$ discrete rotations $\mathbf{u}_l^{(m)}$ for each discrete position $\mathbf{q}_m$. Furthermore, according to the 3GPP standard \cite{3gpp}, the effective antenna gain $A(\tilde{\theta}_{b,\iota,k}, \tilde{\phi}_{b,\iota,k})$ in \eqref{gm} in dBi for a user's signal arrival angles  $(\tilde{\theta}_{b,\iota,k}, \tilde{\phi}_{b,\iota,k})$ at the BS is given by \cite{3gpp}
\begin{align}\label{gm1}
\!A(\tilde{\theta}_{b,\iota,k}, \tilde{\phi}_{b,\iota,k})\!=\!G_{\max}\!-\!\min\!\left\{\!-[A_{\mathrm{H}}(\tilde{\phi}_{b,\iota,k})
\!+\!A_{\mathrm{V}}(\tilde{\theta}_{b,\iota,k})],G_s\!\right\}\!,
\end{align}
with
\begin{align}
A_{\mathrm{H}}(\tilde{\phi}_{b,\iota,k})=-\min\left\{12\left(\frac{\tilde{\phi}_{b,\iota,k}}{\phi_{\mathrm{3dB}}}
\right)^2, G_s\right\},\label{AH}
\end{align}
and
\begin{align}
A_{\mathrm{V}}(\tilde{\theta}_{b,\iota,k})=-\min\left\{12\left(\frac{\tilde{\theta}_{b,\iota,k}}{\theta_{\mathrm{3dB}}}
\right)^2,G_v\right\}, \label{Av}
\end{align}
denoting the horizontal and vertical radiation pattern functions in dBi, respectively, where $\theta_{\mathrm{3dB}}$ and $\phi_{\mathrm{3dB}}$ are the 3-dB beamwidth parameters both set to $65^\circ$, $G_s=25$ dBi denotes the front-back ratio, and $G_v=25$ dBi represents the sidelobe
level limit \cite{3gpp}. In the above, $G_{\max}=8$ dBi represents the peak directional gain of each antenna element in the main lobe direction.

Moreover, we consider the following benchmark schemes for offline optimization, all of which are based on the traditional sector antennas at the BS, with the number of sector antennas equal to three. Each sector antenna is composed of $\lceil\frac{NB}{3}\rceil$ vertically positioned directional antennas, so that the total number of antennas at the BS is (approximately) equal to that of the proposed 6DMA-BS.
\begin{itemize}
\item \textbf{FPA}: This scheme arranges each sector antenna to cover approximately $120^\circ$ while maintaining the positions and rotations of all sector antennas static (with a downtilt of $15^\circ$).

\item \textbf{6DMA with circular discrete positions}: This scheme fixes the downtilts of the three sector antennas as $15^{\circ}$. However, it allows each sector antenna's center to move along a circular path parallel to the ground over $M$ equally-spaced discrete positions.

\item \textbf{6DMA with discrete rotations only}: In this scheme, the positions of the three sector antennas are fixed, but each sector antenna can rotate over $L$ equally-spaced discrete angles.
\end{itemize}
\begin{figure}[t!]
\centering
\setlength{\abovecaptionskip}{0.cm}
\includegraphics[width=3.69in]{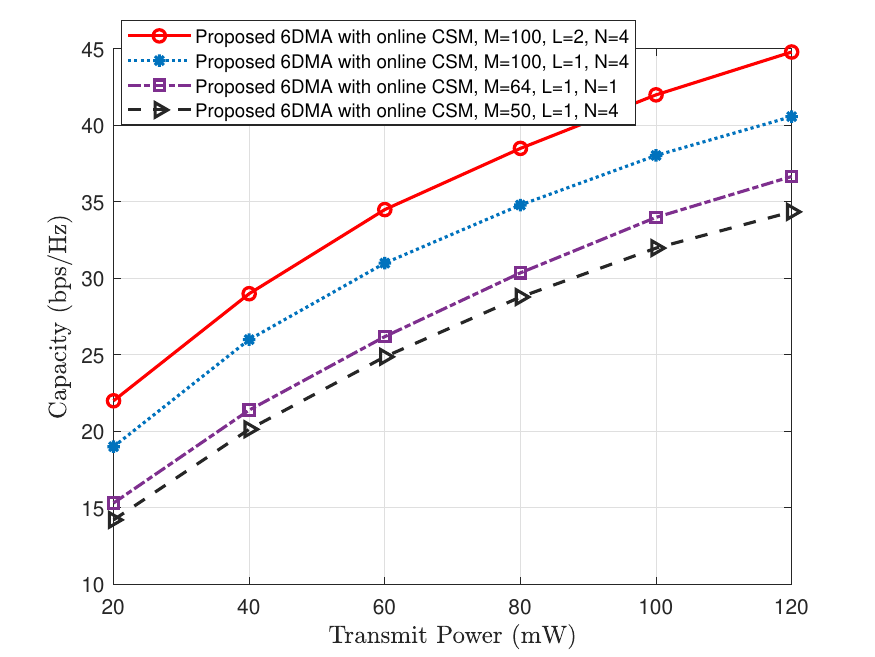}
\caption{Average network capacity versus transmit power for the proposed CSM-based online optimization algorithm.}
\label{CSM}
\end{figure}

First, Fig. \ref{alg_user} shows the average network capacity achieved by the proposed offline optimization algorithm with statistical channel knowledge versus the users' transmit power. It is observed that the network capacity increases as the transmit power increases. Moreover, the network capacity for the proposed offline optimization algorithm increases as the number of discrete positions, $M$, and/or the number of discrete rotations, $L$, increase.
This is expected because with more discrete positions and rotations to choose from, there are more spatial DoFs for tuning the 6DMA surfaces to better adapt to the (non-uniform) spatial distribution of the users, thereby improving the array gains achieved by their directional antennas as well as the spatial multiplexing gains due to more effective multiuser interference mitigation.

In Fig. \ref{CSM}, we evaluate the performance of the proposed CSM-based online algorithm without statistical channel knowledge. It is observed that given the same number of antennas at each 6DMA surface (i.e., $N=4$), for larger values of $M$ and/or $L$, the network capacity increases thanks to the improved flexibility in positioning/rotating the 6DMA surfaces. Interestingly, it is also observed that the performance with $N = 4$ is inferior to that with $N=1$, given the same total number of antennas of all 6DMA surfaces, i.e., $NB=64$. This is because a smaller $N$ leads to a larger $B$ and thus more flexibility for positioning/rotating the $NB$ antennas at the 6DMA-BS. However, larger values of $B$ imply higher implementation cost in terms of movement hardware and power consumption. Thus a balanced trade-off between 6DMA system performance and cost is desirable in practice.

\begin{figure}[!t]
\subfigure[$\xi=0$.]{
\begin{minipage}{8cm}
\centering
\includegraphics[scale=0.61]{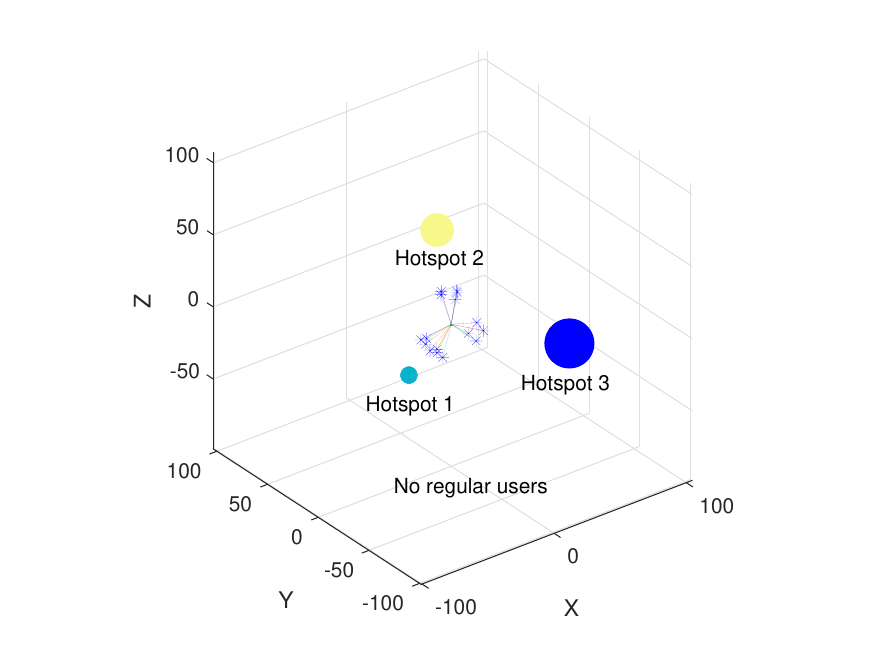}
\end{minipage}
}
\hspace*{-5mm}
\subfigure[$\xi=0.8$.]{
\begin{minipage}{8cm}
\centering
\includegraphics[scale=0.61]{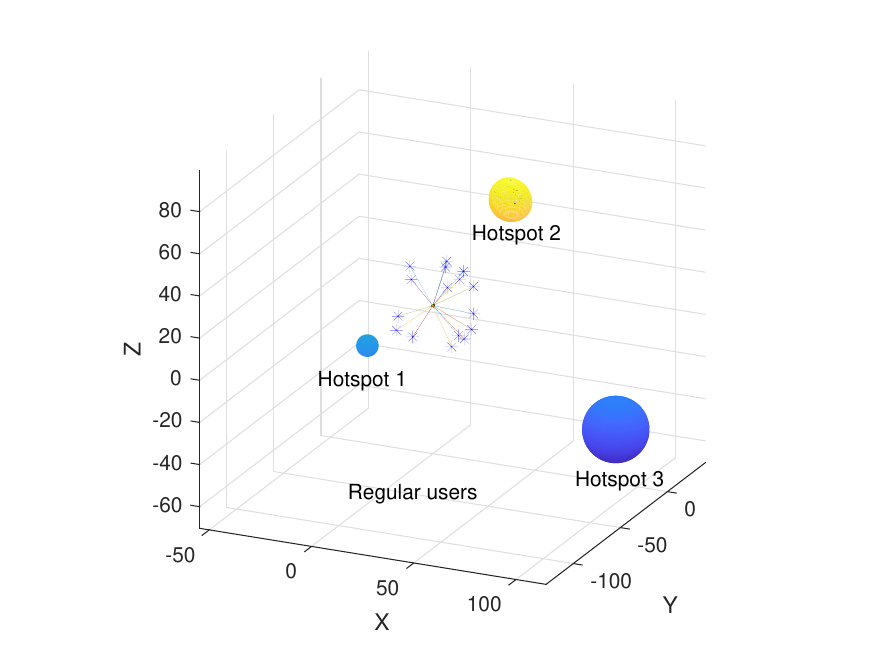}
\end{minipage}
}
\caption{Discrete positions and rotations of 6DMA surfaces optimized via the CSM-based online algorithm for different user spatial distributions (the direction of the snowflake denotes the normal vector of each 6DMA surface).}
\label{USERS}
\end{figure}

Next, in Fig. \ref{USERS}, we illustrate the optimized discrete positions/rotations of the 6DMA surfaces via the proposed CSM-based online algorithm for different  spatial user distributions. As depicted in Fig. \ref{USERS}(a), when the network consists of hotspot users only (i.e., $\xi=0$), the 6DMA surfaces generally align themselves towards the hotspot sub-regions. The distribution of the 6DMA surfaces over different  hotspots is influenced by both their average distances from the BS and the user densities therein.
In comparison, Fig. \ref{USERS}(b) shows the optimized positions/rotations of 6DMA surfaces when the users constitute both hotspot and regular users (with $\xi=0.8$). In this case, it is observed that the 6DMA surfaces are distributed more uniformly in space to cater to both types of users.

\begin{figure}[t!]
\centering
\setlength{\abovecaptionskip}{0.cm}
\includegraphics[width=3.69in]{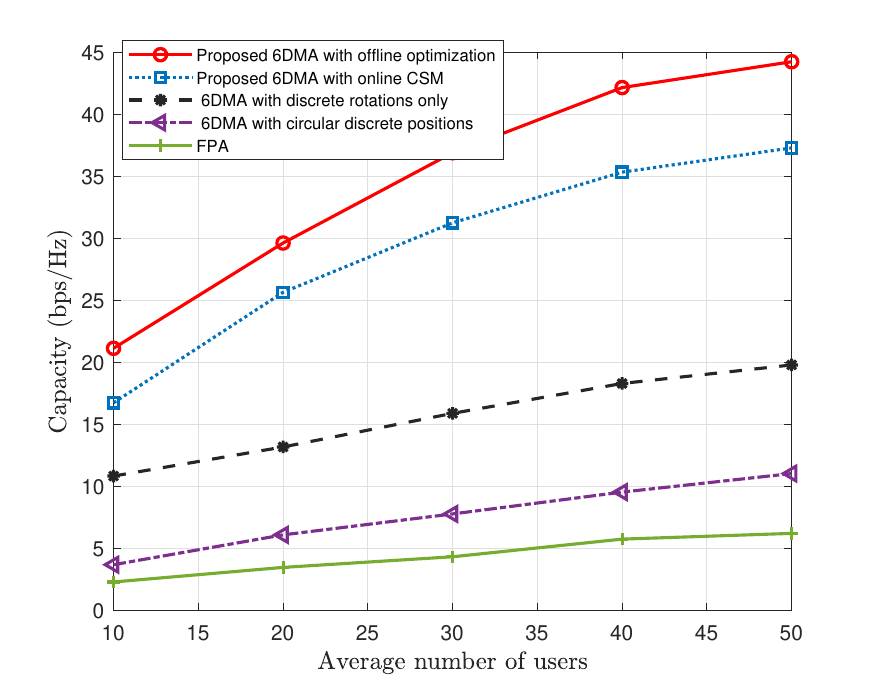}
\caption{Average network capacity versus the average
number of users, $\mu$.}
\label{usernumber}
\end{figure}

In Fig. \ref{usernumber}, we compare the average network capacity for different schemes versus the average number of users, $\mu$, by
setting $M=600$ and $L=27$ for offline optimization, and $M=100$ and $L=2$ for online optimization.
First, it is observed that the proposed offline optimization algorithm performs better than the CSM-based online algorithm, due to the following two reasons. On the one hand, the numbers of discrete positions/rotations (i.e., $M$ and $L$, respectively) in the former case are much larger than their counterparts in the latter case, indicating more flexibility for 6DMA surface movement.
On the other hand, this also demonstrates the benefit of having the statistical channel information in offline optimization, which is not available for CSM-based online optimization. It is worth pointing out that the
offline algorithm has a higher computational complexity than the CSM-based online algorithm, as shown in Sections IV and V.

Moreover, from Fig. \ref{usernumber}, it is observed that the network capacity for the proposed offline and CSM-based online algorithms is significantly higher than that of the three benchmark schemes considered, due to their limited position/rotation adaptability. In addition, the performance gaps between
the proposed offline optimization/CSM-based online algorithm and the benchmark  schemes become larger as the average number of users increases, since the higher flexibility in the antenna positions/rotations helps mitigate multiuser interference, which is particularly important for a large number of users when
the system capacity becomes more interference-limited.

\begin{figure}[t!]
\centering
\setlength{\abovecaptionskip}{0.cm}
\includegraphics[width=3.69in]{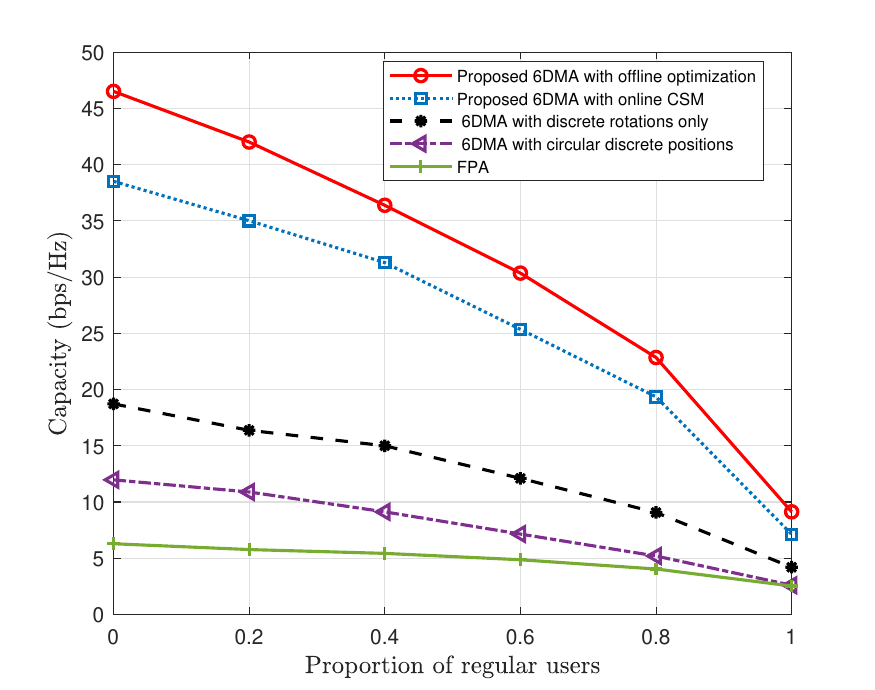}
\caption{Average network capacity versus proportion of regular users, $\xi$.}
\label{back}
\end{figure}

Finally, in Fig. \ref{back}, we plot the average network capacity versus the proportion of regular users, $\xi$. The parameters of both proposed algorithms are set to the same value as for Fig. \ref{usernumber}.
We observe that the performances of the proposed algorithms and benchmark schemes all deteriorate as $\xi$ increases, i.e., the user spatial distribution
becomes uniform. This is because, for uniform user distribution, the exact positions and rotations of the 6DMA surfaces become less useful for performance improvement and the average user-BS distances in the cell become larger.
Furthermore, we observe that the proposed algorithms outperform the benchmark schemes for all values of $\xi$, while the performance gaps increase for smaller $\xi$, i.e., as the user distribution becomes more spatially non-uniform. The above results indicate the more effectiveness of the proposed 6DMA system in exploiting spatially diverse/clustered user distributions.

\section{Conclusion}
In this paper, we considered 6DMA-BS under practical discrete
position and rotation constraints. Specifically, the 3D positions and 3D rotations of the 6DMA surfaces at the BS were jointly optimized for maximization of the average network capacity while meeting the practical discrete movement constraints.
We proposed both offline and online algorithms to solve the optimization problem for the two practical cases when the channel statistics are a priori known and unknown, respectively, by applying the Monte Carlo and CSM methods. Extensive simulation results were presented for various practical setups, which showed that significant network capacity gains are achieved by the
6DMA-BS over benchmark BS
architectures with FPAs and 6DMAs with limited movability. These gains are attributed to the joint selection of the antennas' positions and rotations based on the spatial user distribution, either explicitly (with statistical channel knowledge) or implicitly (without statistical channel knowledge). The results of this paper thus pave a new and practical way to enhance MIMO system capacity without increasing the number of antennas in future-generation wireless networks.

\bibliographystyle{IEEEtran}
\bibliography{fabs}

\begin{thebibliography}{10}
\providecommand{\url}[1]{#1}
\csname url@samestyle\endcsname
\providecommand{\newblock}{\relax}
\providecommand{\bibinfo}[2]{#2}
\providecommand{\BIBentrySTDinterwordspacing}{\spaceskip=0pt\relax}
\providecommand{\BIBentryALTinterwordstretchfactor}{4}
\providecommand{\BIBentryALTinterwordspacing}{\spaceskip=\fontdimen2\font plus
\BIBentryALTinterwordstretchfactor\fontdimen3\font minus
  \fontdimen4\font\relax}
\providecommand{\BIBforeignlanguage}[2]{{%
\expandafter\ifx\csname l@#1\endcsname\relax
\typeout{** WARNING: IEEEtran.bst: No hyphenation pattern has been}%
\typeout{** loaded for the language `#1'. Using the pattern for}%
\typeout{** the default language instead.}%
\else
\language=\csname l@#1\endcsname
\fi
#2}}
\providecommand{\BIBdecl}{\relax}
\BIBdecl

\bibitem{LA10}
E.~G. Larsson, O.~Edfors, F.~Tufvesson, and T.~L. Marzetta, ``Massive {MIMO}
  for next generation wireless systems,'' \emph{IEEE Commun. Mag.}, vol.~52,
  no.~2, pp. 186--195, Feb. 2014.

\bibitem{ruig}
L.~Lu, G.~Y. Li, A.~L. Swindlehurst, A.~Ashikhmin, and R.~Zhang, ``An overview
  of massive {MIMO}: {Benefits} and challenges,'' \emph{IEEE J. Sel. Top.
  Signal Process.}, vol.~8, no.~5, pp. 742--758, Oct. 2014.

\bibitem{free}
H.~Q. Ngo, A.~Ashikhmin, H.~Yang, E.~G. Larsson, and T.~L. Marzetta,
  ``Cell-free massive {MIMO} versus small cells,'' \emph{IEEE Trans. Wireless
  Commun.}, vol.~16, no.~3, pp. 1834--1850, Mar. 2017.

\bibitem{free0}
E.~Nayebi, A.~Ashikhmin, T.~L. Marzetta, H.~Yang, and B.~D. Rao, ``Precoding
  and power optimization in cell-free massive {MIMO} systems,'' \emph{IEEE
  Trans. Wireless Commun.}, vol.~16, no.~7, pp. 4445--4459, Jul. 2017.

\bibitem{exl}
Z.~Wang, J.~Zhang, H.~Du, D.~Niyato, S.~Cui, B.~Ai, M.~Debbah, K.~B. Letaief,
  and H.~V. Poor, ``A tutorial on extremely large-scale {MIMO} for {6G}:
  {Fundamentals}, signal processing, and applications,'' \emph{IEEE Commun.
  Surv. Tutorials.}, vol.~PP, no.~99, pp. 1--1, Jan. 2024.

\bibitem{zeng}
H.~Lu and Y.~Zeng, ``Communicating with extremely large-scale array/surface:
  {Unified} modeling and performance analysis,'' \emph{IEEE Trans. Wireless
  Commun.}, vol.~21, no.~6, pp. 4039--4053, Jun. 2022.

\bibitem{first}
X.~Shao, Q.~Jiang, and R.~Zhang, ``{6D} movable antenna based on user
  distribution: {Modeling} and optimization,''
  \emph{https://arxiv.org/abs/2403.08123}, Mar. 2024.

\bibitem{9388928}
W.~K. New, K.-K. Wong, H.~Xu, K.-F. Tong, C.-B. Chae, and Y.~Zhang, ``Fluid
  antenna system enhancing orthogonal and non-orthogonal multiple access,''
  \emph{IEEE Commun. Lett.}, vol.~28, no.~1, pp. 218--222, Jan. 2024.

\bibitem{9264694}
K.-K. Wong, A.~Shojaeifard, K.-F. Tong, and Y.~Zhang, ``Fluid antenna
  systems,'' \emph{IEEE Trans. Wireless Commun.}, vol.~20, no.~3, pp.
  1950--1962, Mar. 2021.

\bibitem{zhu2}
L.~Zhu, W.~Ma, B.~Ning, and R.~Zhang, ``Movable-antenna enhanced multiuser
  communication via antenna position optimization,'' \emph{IEEE Trans. Wireless
  Commun.}, pp. 1--1, Dec. 2023.

\bibitem{ma}
W.~Ma, L.~Zhu, and R.~Zhang, ``Compressed sensing based channel estimation for
  movable antenna communications,'' \emph{IEEE Commun. Lett.}, vol.~27, no.~10,
  pp. 2747--2751, Oct. 2023.

\bibitem{yifei}
Y.~Wu, D.~Xu, D.~W.~K. Ng, W.~Gerstacker, and R.~Schober, ``Movable
  antenna-enhanced multiuser communication: Jointly optimal discrete antenna
  positioning and beamforming,'' in \emph{IEEE Global Commun. Conf.
  (GLOBECOM)}, Dec. 2023, pp. 7508--7513.

\bibitem{linear}
D.~P. Bertsekas, ``Nonlinear programming,'' \emph{J. Oper. Res. Soc.}, vol.~48,
  no.~3, pp. 334--334, 1997.

\bibitem{mk}
D.~Liu, K.~Sun, Z.~Wang, R.~Liu, and Z.-J. Zha, ``Frank-wolfe network: An
  interpretable deep structure for non-sparse coding,'' \emph{IEEE Trans.
  Circuits Syst. Video Technol.}, vol.~30, no.~9, pp. 3068--3080, Sep. 2020.

\bibitem{trip}
B.~Keinert, M.~Innmann, M.~S{\"a}nger, and M.~Stamminger, ``Spherical
  {Fibonacci} mapping,'' \emph{ACM Trans. Graphics (TOG)}, vol.~34, no.~6, pp.
  1--7, 2015.

\bibitem{proc}
Q.~Wu \emph{et~al.}, ``Intelligent surfaces empowered wireless network: Recent
  advances and the road to {6G},'' \emph{arXiv preprint arXiv:2312.16918},
  2023.

\bibitem{shuyi}
S.~Ren, K.~Shen, Y.~Zhang, X.~Li, X.~Chen, and Z.-Q. Luo, ``Configuring
  intelligent reflecting surface with performance guarantees: Blind
  beamforming,'' \emph{IEEE Trans. Wireless Commun.}, vol.~22, no.~5, pp.
  3355--3370, May 2023.

\bibitem{shaoirs}
X.~Shao, L.~Cheng, X.~Chen, C.~Huang, and D.~W.~K. Ng, ``Reconfigurable
  intelligent surface-aided {6G} massive access: {Coupled} tensor modeling and
  sparse bayesian learning,'' \emph{IEEE Trans. Wireless Commun.}, vol.~21,
  no.~12, pp. 10\,145--10\,161, Dec. 2022.

\bibitem{rot3}
J.~Diebel \emph{et~al.}, ``Representing attitude: {Euler} angles, unit
  quaternions, and rotation vectors,'' \emph{Matrix}, vol.~58, no. 15-16, pp.
  1--35, 2006.

\bibitem{as9}
S.~Sanayei and A.~Nosratinia, ``Antenna selection in {MIMO} systems,''
  \emph{IEEE Commun. Magazine}, vol.~42, no.~10, pp. 68--73, Oct. 2004.

\bibitem{as10}
A.~Molisch and M.~Win, ``{MIMO} systems with antenna selection,'' \emph{IEEE
  Microwave Mag.}, vol.~5, no.~1, pp. 46--56, Mar. 2004.

\bibitem{sele}
Y.~Gao, H.~Vinck, and T.~Kaiser, ``Massive {MIMO} antenna selection:
  {Switching} architectures, capacity bounds, and optimal antenna selection
  algorithms,'' \emph{IEEE Trans. Signal Process.}, vol.~66, no.~5, pp.
  1346--1360, Mar. 2018.

\bibitem{dis0}
F.~Glover and E.~Woolsey, ``Converting the 0-1 polynomial programming problem
  to a 0-1 linear program,'' \emph{Operations {Research}}, vol.~22, no.~1, pp.
  180--182, 1974.

\bibitem{nonl1}
X.~Gao, O.~Edfors, J.~Liu, and F.~Tufvesson, ``Antenna selection in measured
  massive {MIMO} channels using convex optimization,'' in \emph{EEE Globecom
  Workshops (GC Wkshps)}, Dec 2013, pp. 129--134.

\bibitem{lin}
T.~Rocha, A.~Borges, S.~Paredes, and A.~Pinho, ``A {Matlab} tool for solving
  linear goal programming problems,'' in \emph{Experiment Int. Conf.}, Jun.
  2019, pp. 337--342.

\bibitem{flow1}
K.~Son, H.~Kim, Y.~Yi, and B.~Krishnamachari, ``Base station operation and user
  association mechanisms for energy-delay tradeoffs in green cellular
  networks,'' \emph{IEEE J. Sel. Areas Commun.,}, vol.~29, no.~8, pp.
  1525--1536, Sep. 2011.

\bibitem{cluter}
C.~Saha, H.~S. Dhillon, N.~Miyoshi, and J.~G. Andrews, ``Unified analysis of
  {HetNets} using {Poisson} cluster processes under max-power association,''
  \emph{IEEE Trans. Wireless Commun.}, vol.~18, no.~8, pp. 3797--3812, Aug.
  2019.

\bibitem{3gpp}
G.~T. 36.873:, ``Study on {3D} channel model for {LTE},'' vol. V12.7.0, Dec.
  2017.

\end{thebibliography}
\end{document}